# VISCOSITIES OF IODOBENZENE + *n*-ALKANE MIXTURES AT (288.15-308.15) K. MEASUREMENTS AND RESULTS FROM MODELS

Luis Felipe Sanz[a], Juan Antonio González,[a*] Fernando Hevia[a], Daniel Lozano-Martín[a], Isaías García de la Fuente,[a] José Carlos Cobos[a]

[a]G.E.T.E.F., Departamento de Física Aplicada, Facultad de Ciencias, Universidad de Valladolid, Paseo de Belén, 7, 47011 Valladolid, Spain.

*corresponding author, e-mail: jagl@termo.uva.es; Fax: +34-983-423136; Tel: +34-983-423757

**Abstract**

Kinematic viscosities were measured for iodobenzene + $n$-alkane mixtures at (288.15-308.15) K and atmospheric pressure. The corresponding dynamic viscosities ($\eta$) were also determined using density data previously obtained in our laboratory. This set of data was employed to calculate $\Delta \eta$ (deviations in absolute viscosity) and magnitudes of viscous flow. In addition, the correlation equations: McAllister, Grunberg-Nissan, Fang-He, and the Bloomfield-Dewan´s model were applied to the systems: iodobenzene, or 1-chloronaphthalene, or 1,2,4-trichlorobenzene, or methyl benzoate or benzene or cyclohexane + $n$-alkane. It is remarkable that, within the Bloomfield-Dewan´s model, residual Gibbs energies were calculated using DISQUAC with interaction parameters available in the literature. From the dependence of $U_{V_m}^E$ (isochoric molar excess internal energy) and $\Delta \eta$ with $n$ (the number of C atoms of the $n$-alkane), it is shown that the loss of fluidization of mixtures containing iodobenzene, 1,2,4-trichlorobenzene, or 1-chloronaphthalene when $n$ increases can be ascribed to a decrease upon mixing of the number of broken interactions between like molecules. The breaking of correlations of molecular orientations characteristic of longer $n$-alkanes may explain the decreased negative $\Delta \eta$ values of benzene mixtures with $n$ =14,16. The replacement, in this type of systems, of benzene by cyclohexane, leads to increased positive $\Delta \eta$ values, probably due to the different shape of cyclohexane. On the other hand, binary mixtures formed by an aromatic polar compound mentioned above and a short $n$-alkane show large structural effects and large negative $\Delta \eta$ values.

From the application of the models, it seems that dispersive interactions are dominant and that size effects are not relevant on $\eta$ values. The free volume model provides good results for most of the systems considered, since deviations are less than 6% for 20 mixtures from the 29 solutions under study, and only 4 systems show deviations higher than 10%, with a maximum deviation of 15%. Results improve when, within the Bloomfield-Dewan's theory, the contribution to $\eta$ of the absolute reaction rate model is also considered.

## 1. Introduction

We are engaged in a systematic investigation on mixtures formed by a molecule A, of plate-like or of more or less globular shape, and *n*-alkane since in such solutions two interesting effects may exist when long chain *n*-alkanes are involved: the Patterson's or the Wilhelm's effects [1-4]. The former is typically encountered when A, a quasispherical molecule (benzene, cyclohexane, $CCl_4$), is mixed with a long *n*-alkane. In this case, there is an extra endothermic contribution to $H_m^E$, the excess molar enthalpy, due to the breaking of the local order characteristic of longer *n*-alkanes (correlations of molecular orientations, CMO) [1,2,5,6]. The Wilhelm's effect is encountered in systems including a flat compound, e.g., 1,2,4-trimethylbenzene [7] or 1,2.4-trichlorobenzene [4] or 1-chloronaphthalene [8]. These solutions show decreasing values of $H_m^E$ at equimolar composition when *n*, the number of C atoms of the *n*-alkane, is increased. This behaviour has been explained by assuming the creation of some type of intramolecular order due to the flat component hinders the rotational motion of the segments of the flexible molecules of longer *n*-alkanes. In the recent past, we have focused our work on binary systems formed by an *n*-alkane and fluorobenzene [9], or chlorobenzene, or bromobenzene, or 1,2,4-trichlorobenzene, or 1-chloronaphthalene [10], or a bicyclic compound such as tetralin, bicyclohexyl, cyclohexylbenzene, or decalin [11]. Briefly, the main conclusions of these investigations may be summarized as follows. (i) Dispersive interactions are dominant. (ii) Solutions with shorter *n*-alkanes show large structural effects. (iii) It has been demonstrated that excess molar properties at constant volume, internal energies ($U_{Vm}^E$) and heat capacities, are very useful tools to attain a better understanding of interactional and structural effects present in the systems. (iv) Two competing contributions to $U_{Vm}^E$ exist: (a) a poorer ability of longer *n*-alkanes to break interactions between A molecules, which leads to decreased values of $U_{Vm}^E$, and (b) an extra endothermic contribution to this excess functions which arises from the breaking of CMO of longer *n*-alkanes. If the first contribution is dominant, then $U_{Vm}^E$ decreases when *n* is increased. This is the case of mixtures with chlorobenzene, bromobenzene, 1,2,4-trichlorobenzene, or 1-chloronaphthalene and it is also expected in solutions with iodobenzene. Systems involving cyclohexane or benzene behave differently: $U_{Vm}^E$ decreases up to *n* = 8 and from $n \geq 10$ increases, which indicates that, for the solutions with longer *n*-alkanes, the second contribution to $U_{Vm}^E$ is dominant.

As continuation of these works, we have recently provided excess molar volumes, $V_m^E$, for iodobenzene + heptane, or + decane, or + dodecane, or + tetradecane mixtures over the temperature range (288.15-308.15) K [12] and now we report viscosities for the same solutions over the same range of temperature. Some viscosity data are available in the literature for binary mixtures containing different *n*-alkanes and chlorobenzene [13,14], or bromobenzene [14], or

1,2,4-trichlorobenzene [15], or 1-chloronaphthalene [16]. In addition, the new data are correlated by means of different equations: McAllister [17], Grunberg-Nissan [18], Fang-He [19], and also using the model proposed by Bloomfield and Dewan [20], which contains contributions to dynamic viscosity from the free volume and absolute reaction rate theories. For the sake of completeness, the application of these models/equations is extended to other systems containing *n*-alkane and an aromatic molecule (benzene, 1-chloronaphthalene, 1,2,4-trichlorobenzene, 1-chloronaphthalene, methyl benzoate), or cyclohexane. The present measurements, together with those reported previously for density [12], are also used to calculate molar magnitudes of activation of viscous flow: Gibbs energy, $\Delta G_m^*$, enthalpy, $\Delta H_m^*$, and entropy, $\Delta S_m^*$ from the application of the Eyring's theory [21-23].

## 2. Experimental

Information on the source and purity of the chemicals is included in Table 1. Solutions were prepared by weighing in small vessels of about 10 cm$^3$. The concentration of the mixtures (given by the mole fraction of iodobenzene) was calculated from mass measurements. The masses were determined by weighing them using an analytical balance (MSU125P, Sartorius) and correcting for buoyancy effects, with a standard uncertainty of $5 \cdot 10^{-5}$ g. Along the process, caution was taken in order to prevent evaporation. Conversion to molar quantities was based on the relative atomic mass Table of 2015 issued by I.U.P.A.C [24]. The error in the final mole fraction is estimated to be 0.0010. All measurements were carried out at atmospheric pressure.

Kinematic viscosities, $\nu$, were determined using an Ubbelohde viscosimeter with a Schott-Geräte automatic measuring unit model AVS-350. The temperature was hold constant within $\pm$ 0.02 K by means of a controller bath CT52, also from Schott. Details on the calibration of the apparatus can be found elsewhere [25]. Values of dynamic viscosity ($\eta$) were obtained using densities determined early [12]. The uncertainties of the kinematic and dynamic viscosities are estimated at $\pm$ 1% and 1.1%, respectively. Table 2 shows that there is a rather good agreement between values of $\nu$ and $\eta$ of the pure compounds with results available in the literature. The mean deviations for the $\eta$ values of *n*-alkanes at any temperature are: 1.1% (heptane); 0.5% (decane); 0.7% (dodecane); 0.9% (tetradecane), with a maximum deviation of 2%. For kinematic viscosities, the mean deviations are similar. For iodobenzene, we have found in the literature very few $\eta$ data and the mean deviation is somewhat poorer: 2.9%. The uncertainty of the deviations of dynamic viscosity from linear dependence on molar fraction (hereafter, deviations in absolute viscosity, see below) does not exceed from $\pm$ 2.5%. The comparison of our results with those available in the literature for the 1-propanol + dimethyl carbonate mixture system at (288.15-313.15) K [26] supports such statement.

### 3. Models

*3.1 Absolute reaction rate model*

This theory, developed by Eyring and co-workers [21-23], relates viscosity to the free energy required for a molecule to flow from an equilibrium position to a new one, overcoming the attractive interactions caused by its neighbours. The expression for dynamic viscosity is:

$$\eta = \frac{hN_A}{V_m} \exp[\frac{\Delta G_m^*}{RT}] \quad (1)$$

where $h$ is the Planck's constant, $N_A$, the Avogadro's number and $V_m$ the molar volume. The values of $\Delta G_m^* (= \Delta H_m^* - T\Delta S_m^*)$ were calculated determining previously the required values of $\Delta H_m^*$ and $\Delta S_m^*$ from the plots $\ln \frac{\eta V_m}{hN_A}$ vs. $1/T$ [27,28]. These ones give a straight line for each mixture and $\Delta H_m^*$, and $\Delta S_m^*$ can be estimated from its slope and intercept.

*3.2 Kinematic viscosity*

Results of kinematic viscosities have been correlated using the McAllister equation, based on the Eyring's theory [21] and a three-body interaction model [17]:

$$\ln \nu = x_1^3 \ln \nu_1 + 3x_1^2 x_2 \ln Z_{12} + 3x_1 x_2^2 \ln Z_{21} + x_2^3 \ln \nu_2 - \ln(x_1 + x_2 \frac{M_2}{M_1})$$
$$+ 3x_1^2 x_2 \ln(\frac{2}{3} + \frac{M_2}{3M_1}) + 3x_1 x_2^2 \ln(\frac{1}{3} + \frac{2M_2}{3M_1}) + x_2^3 \ln(\frac{M_2}{M_1}) \quad (2)$$

where $M_i$ and $\nu_i$ are, respectively, the molar mass and the kinematic viscosity of component i, and $Z_{12}, Z_{21}$ the adjustable parameters.

*3.3 Dynamic viscosity: equations with one adjustable parameter*

Values of dynamic viscosity were correlated using the Grunberg-Nissan [18] and the Fang-He [19] equations. According to the former equation, viscosity is calculated from the expression [18]:

$$\eta = \exp[(x_1 \ln \eta_1 + x_2 \ln \eta_2 + x_1 x_2 G_{12})] \quad (3)$$

where $\eta_i$ stands for the dynamic viscosity of component i. The Fang-He equation combines the Eyring's model with a modified Flory-Huggins equation expressed in terms of surface fractions [19]:

$$\ln \eta = (\theta_1 \ln \eta_1 + \theta_2 \ln \eta_2) + (x_1 \ln \frac{\theta_1}{x_1} + x_2 \ln \frac{\theta_2}{x_2}) + (x_1 V_{m1}^{2/3} + x_2 V_{m2}^{2/3})(W_{12}/RT)\theta_1\theta_2 \quad (4)$$

where $\theta_i$ is the molecular surface fraction defined by $\theta_i = \dfrac{x_i V_{mi}^{2/3}}{\sum x_j V_{mj}^{2/3}}$, being $V_{mi}$ the molar volume of component i). In equations (3) and (4), the adjustable parameters are, respectively, $G_{12}$ and $W_{12}$

*3.4 Bloomfield-Dewan's model*

This theory [20] combines the absolute reaction rate model with the free volume theory [29]. The latter relates viscosity to the probability of occurrence of an empty neighboring site where a molecule can jump. Thus, dynamic viscosity can be determined from the equation:

$$\ln \eta = (x_1 \ln \eta_1 + x_2 \ln \eta_2) + \alpha \ln \eta_{fv} + \beta \ln \eta_{ar} \qquad (5)$$

This means that the probability for viscous flow is calculated as the product of the probabilities of having the sufficient activation energy and of the existence of an empty site [20,30]. The parameters $\alpha, \beta$ are weighting factors with values between 0 and 1. In equation (5), $\ln \eta_{fv}$ arises from free volume effects and it is obtained from the expression:

$$\ln \eta_{fv} = \dfrac{1}{\hat{V}_m - 1} - \dfrac{x_1}{\hat{V}_{m1} - 1} - \dfrac{x_2}{\hat{V}_{m2} - 1} \qquad (6)$$

Here, $\hat{V}_m$ and $\hat{V}_{mi}$ are the reduced volumes of the mixture, and of component i, respectively, defined as in the Flory model [31] ($\hat{V}_{mi} = V_{mi}/V_{mi}^*$, being $V_{mi}^*$ the reduction volume). In addition,

$$\hat{V}_m = \dfrac{V_m^E}{x_1 V_{m1}^* + x_2 V_{m2}^*} + \Psi_1 \hat{V}_1^* + \Psi_2 \hat{V}_2^* \qquad (7)$$

and $\Psi_i \left(= \dfrac{x_i V_{mi}^*}{\sum x_j V_{mj}^*}\right)$ is the segment fraction of component i. The contribution arising from the absolute reaction rate model is given by:

$$\ln \eta_{ar} = -\dfrac{\Delta G_m^{RES}}{RT} \qquad (8)$$

Typically, the values of $\Delta G_m^{RES}$ are obtained from the application of the Flory model using the corresponding expressions for $H_m^E$ and for the residual entropy [20]. In this work, we have applied a different approach and $\Delta G_m^{RES}$ values have been calculated using the DISQUAC model [32] with interaction parameters previously determined [10,33-35]. Essentially, $\Delta G_m^{RES}$ is obtained from the excess Gibbs energy, $G_m^E$, by subtracting the corresponding combinatorial term represented by the Flory-Huggins equation.

## 4. Results

### 4.1 Experimental results

Table 3 lists the experimental values of $\nu$ and $\eta$ determined in this work for $C_6H_5I$ + $n$-alkane mixtures *vs.* $x_1$, the mole fraction of the polar component. We have also determined deviations in absolute viscosity (a non-Gibbsian property, see Table 3) according to the equation:

$$\Delta \eta = \eta - (x_1\eta_1 + x_2\eta_2) \qquad (9)$$

Results at 298.15 K are shown graphically in Figure 1. Values of $\Delta G_m^*$, $\Delta H_m^*$, $\Delta S_m^*$ and of $\Delta(\Delta G_m^*) = \Delta G_m^* - x_1\Delta G_{m1}^* - x_2\Delta G_{m2}^*$ are collected in Table S1 of supplementary material (see Figure 2). Data for $\Delta \eta$ and $\Delta(\Delta G_m^*)$ were fitted by unweighted least-squares polynomial regression to the equation:

$$Q = x_1(1-x_1)\sum_{i=0}^{k-1} A_i (2x_1 - 1)^i \qquad (10)$$

being $Q = \Delta \eta, \Delta(\Delta G_m^*)$. The number of coefficients $k$ used in equation (10) for each mixture was determined by applying an F-test [36] at the 99.5 % confidence level. Table 4 lists the parameters $A_i$ obtained in the regressions, together with the standard deviations, $\sigma(Q)$, defined by:

$$\sigma(Q) = \left[ \frac{1}{N-k} \sum (Q_{cal} - Q_{exp})^2 \right]^{1/2} \qquad (11)$$

where $N$ is the number of direct experimental values.

For viscosity, results obtained from the application of the different models/equations considered in the present study are collected in Tables 5-6 and Table S2 (Figures 3-8), which also list the adjusted parameters: $Z_{12}, Z_{21}$ (equation (2)) (Table 5), $G_{12}$ (equation (3), Tables 5 and S2), $W_{12}/RT$ (equation (4), Table S2), as well as the values of $\alpha$ and $\beta$ (equation (5), Table 6). Particularly, results of the correlations are compared by means of the relative standard deviations, $\sigma_r(F)$, calculated from:

$$\sigma_r(F) = \left[ \frac{1}{N} \sum \left( \frac{F_{cal} - F_{exp}}{F_{exp}} \right)^2 \right]^{1/2} \qquad (12)$$

where, $F = \eta$ or $\nu$.

## 5. Discussion

Hereafter, thermophysical properties are considered at equimolar composition and 298.15 K.

*5.1 Viscosity results*

Firstly, we note that the $\eta(x_1)$ curves for iodobenzene + dodecane, or + tetradecane systems show a minimum (Table 3). The same occurs in the mixtures: 1,2-dibromethane + *m*-xylene, or + *p*-xylene, or *o*-xylene [37], or ethanol + benzene [38], or + cyclohexane [39], or in methyl benzoate + dodecane, or + tetradecane [40], or butyl benzoate + hexadecane [41], or in chlorobenzene + decane [13], or in systems with ionic liquids [42]. On the other hand, the curves of kinematic viscosity vs. concentration of the iodobenzene + decane (this work) or of toluene + octane [43] mixtures also show a similar minimum. Therefore, such concentration dependence of the viscosity is encountered in solutions characterized by dispersive or dipolar interactions or containing a self-associated compound. One can conclude then that the existence of the mentioned minima is due to some type of structural effect. Note that the $\eta(x_1)$ curves show a maximum when strong interactions exist between the mixture compounds, an effect that leads to a lower fluidization of the system [25, 44-47]. In addition, viscosity decreases when temperature increases (Table 3), indicating that the mentioned fluidization also increases under such condition. This is the normal behavior [13,15,48-50].

The values of $\Delta \eta$ of iodobenzene + *n*-alkane mixtures are negative (Table 3, Figure 1), a typical feature of systems where no specific interactions exist between the components [51,52]. This occurs, e.g., in 1-alkanol [53-55], or 2-butanone [56] or dialkyl carbonate [57], or methyl alkanoate [58], or 1-chloronaphthalene [16], or methyl benzoate [40], or benzene [59], or cyclohexane [60,61] + *n*-alkane systems. Positive values of $\Delta \eta$ are found in mixtures where strong interactions exist between unlike molecules, such as in the solutions methanol + cyclohexylamine [45], or + 1-propylamine [47], or acetone + chloroform [44] or 3-dimethylamino-1-propylamine + water [46]. It is remarkable that the cyclohexane + tetradecane, or + hexadecane mixtures also show positive values of $\Delta \eta$ [60,61] (see below). This magnitude increases in line with *n* in solutions including iodobenzene (this work, Figure 9), and the same occurs for mixtures with 1,2,4-trichlorobenzene [15], or 1-chloronaphthalene [16], or methyl benzoate [40], or cyclohexane [60,61] (Figure 9). Thus, the increase of the length of the alkane leads to a loss of the fluidization of the system. The variation of $\Delta \eta$ with *n* for systems with benzene is somewhat different: increases smoothly up to $n \geq 10$ and then decreases [59] (Figure 9). To evaluate the influence of interactional effects on the variation of $\Delta \eta$ with *n*, we now consider the corresponding $U_{Vm}^{E}$ values previously determined for most of studied mixtures [10, 11] (Figure 10). Inspection of Figures 9 and 10 suggests that there is a rather close relationship between the variations of $\Delta \eta$ and $U_{Vm}^{E}$ with *n* for mixtures containing iodobenzene, 1,2,4-

trichlorobenzene, or 1-chloronaphthalene where the loss of fluidization of the systems when $n$ increases can be ascribed to a decrease of the number of broken interactions between like molecules upon mixing. Note that $U_{V\text{m}}^{\text{E}}$ decreases with the increasing of $n$ in mixtures with chlorobenzene, bromobenzene, 1,2,4-trichlorobenzene, 1-chloronaphthalene (Figure 10). Similar behavior can be expected for iodobenzene solutions. In the case of systems with methyl benzoate, preliminary estimations provide the following values of $U_{V\text{m}}^{\text{E}}$/J mol$^{-1}$ =1528 ($n$ =6); 1440 ($n$ =7); 1462 ($n$ =8); 1621 ($n$ =12); 1672 ($n$ =14), which is a similar variation to that encountered for mixtures with benzene or cyclohexane [11]. For benzene systems, the observed decrease of $\Delta\eta$ when longer $n$-alkanes are involved may be explained in terms of the corresponding increase of $U_{V\text{m}}^{\text{E}}$, which arises from the disruption of the CMO of such alkanes and that leads to larger negative values of $\Delta\eta$ for $n$ =14,16 (Figure 9). This may indicate that benzene is a good breaker of the CMO characteristic of longer $n$-alkanes. In contrast, systems with cyclohexane show values of $U_{V\text{m}}^{\text{E}}$ and $\Delta\eta$ which increase in line from $n > 10$. The slightly positive $\Delta\eta$ results of mixtures with $n = 14,16$ are noticeable and might be related to the different shape of cyclohexane and that the contribution to $\Delta\eta$ from the disruption of the mentioned CMO is here no so important. At this regards, it is important to underline that $TS_{\text{m}}^{\text{E}}$ ($= H_{\text{m}}^{\text{E}} - G_{\text{m}}^{\text{E}}$) values of mixtures with benzene are much higher than those of solutions with cyclohexane, which can also explain the higher fluidization of benzene systems with $n$ =14,16. For the sake of clarity, some experimental values of $TS_{\text{m}}^{\text{E}}$/J mol$^{-1}$ follow: 501 (931 [62] – 430 [63]; benzene + heptane); 824 (1101 [64] – 188 [65]; benzene + dodecane); 1178 (1256 [64] – 78 [66] benzene + hexadecane); 200 (243 [67] – 43 [68]; cyclohexane + heptane); 425 (360 [69] – (–65) [70]; cyclohexane + dodecane); 679 (501 [71] – (–178) [68]; cyclohexane + hexadecane).

It has been previously stated that a certain correlation exists between $\Delta\eta$ and $V_{\text{m}}^{\text{E}}$, these thermophysical functions having opposite signs [56,72]. According to the results included in Table 6, this is valid for mixtures with iodobenzene ($n > 7$), or methyl benzoate ($n \geq 10$) or 1,2,4-trichlorobenzene ($n > 10$), or cyclohexane ($n \leq 12$) or benzene, since such solutions are characterized by $V_{\text{m}}^{\text{E}} > 0$ and $\Delta\eta < 0$. However, other mixtures show negative values of $\Delta\eta$ and $V_{\text{m}}^{\text{E}}$ (Table 6): iodobenzene + heptane, methyl benzoate + hexane, or + octane, or 1-chloronaphthalene + $n$-alkane; while the systems cyclohexane + tetradecane, or + hexadecane show positive values of $V_{\text{m}}^{\text{E}}$ and $\Delta\eta$ (Table 6). It seems that there is no clear trend at this regards for the homologous series under consideration. On the other hand, both $\Delta\eta$ and $V_{\text{m}}^{\text{E}}$ increase in line with $n$ for systems containing iodobenzene, or 1,2,4-trichlorobenzene, or 1-

chloronaphthalene, or methyl benzoate, or cyclohexane (Table 6, Figures 9 and 11). Solutions with one of the aromatic compounds listed above and a short *n*-alkanes show large structural effects, as indicated by their large negative $V_m^E$ values, and by a higher fluidization (larger negative $\Delta\eta$ results). In such systems, the mentioned effects are of free volume type, and become weaker when *n* increases [10,12] (Figure S1). Since free volume effects lead to increased $\eta$ values [29], one could expect no so negative experimental results of $\Delta\eta$. This suggests that free volume effects do not contribute meaningfully to $\Delta\eta$, a magnitude currently determined, in large extent, by interactional effects (see above). Finally, we underline that the $V_m^E$ values of the systems benzene + tetradecane (1.015 cm$^3$ mol$^{-1}$[59]) or + hexadecane (1.023 cm$^3$ mol$^{-1}$ [73]) are large and positive, which remarks the existence of the breaking of CMO of the alkanes involved in these mixtures, and supports their negative $\Delta\eta$ values.

Systems characterized by interactions between like molecules show positive values of $\frac{\Delta(\Delta\eta)}{\Delta T}$ as in the mixtures ethanol + heptane (4.4 10$^{-3}$ mPa K$^{-1}$) [53], or methyl ester + *n*-alkane (7.6 10$^{-3}$ mPa K$^{-1}$ for methyl pentanoate + octane) [58]. That is, viscosity values approach to those given by ($x_1\eta_1 + x_2\eta_2$), and this means that there is a lower change of the mixture fluidization when *T* increases. For mixtures with idobenzene, we have $\frac{\Delta(\Delta\eta)}{\Delta T}$/ mPa K$^{-1}$= 5.3 10$^{-3}$ (*n* = 7); 4.3 10$^{-3}$ (*n* = 10); 4.8 10$^{-3}$ (*n* = 12); 5.6 10$^{-3}$ (*n* = 14). Negative values of $\frac{\Delta(\Delta\eta)}{\Delta T}$ are encountered for systems where strong interaction between unlike molecules are dominant ($-0.021$ mPa K$^{-1}$ for the methanol + cyclohexylamine mixture [45]), indicating that the mentioned interactions are broken when *T* increases.

*5.2 Results from models*

*5.2.1 Eyring's model*

Firstly, it is remarkable that the current results on $\Delta(\Delta G_m^*)$ (Table S1, Figure 2) are correctly correlated using Redlich-Kister expansions (Table 4), and this means that our measurements at different temperatures are well performed. On the other hand, values of $\Delta G_m^*(x_1)$ show that the change of a molecule from an equilibrium position to a new one overcoming the attractive forces exerted by its neighbours is a process that depends on both enthalpy and entropic effects, since $T\Delta S_m^*$ is nearly $\frac{1}{3}\Delta G_m^*$ (Table S1). In addition, at $x_1 = 0.5$, $\Delta G_m^*$ changes linearly with *n* according to $\Delta G_m^* = 10.15 + 0.457\ n$ ($r = 0.9998$). This remarks the importance of size effects on the relative variation of $\Delta G_m^*$ (*n*). For the sake of comparison, and due to the lack of the required data from the same source to follow the procedure applied in this work to calculate $\Delta G_m^*$, we have estimated this magnitude at equimolar composition and 298.15 K for the other systems under study using directly the equation (1) (Table S3). This allows calculate the corresponding values of $\Delta(\Delta G_m^*)$ (Table S3), since $\Delta G_{m1}^*$ and $\Delta G_{m2}^*$ can be determined similarly. Results show that $\Delta G_m^*$ also changes linearly with *n* (Figure S2), and that, for mixtures with a given alkane, the attractive forces mentioned above become weaker in the sequence: 1-chloronaphthalene > iodobenzene ≈ methylbenzoate > cyclohexane > benzene. Particularly, for decane systems, $\Delta G_m^*$/kJ mol$^{-1}$: 15.7 (1-chloronaphthalene); 15.2 (1,2,4-trichlorobenzene, $T$ = 293.15 K); 14.9 (methyl benzoate); 14.7 (iodobenzene); 14.3 (cyclohexane); 13.6 (benzene). For the mixtures including iodobenzene, the values of $\Delta(\Delta G_m^*)$ also increase with *n*, being negative up to $n$ = 12, and positive for $n$ = 14. Positive values of $\Delta(\Delta G_m^*)$ are also encountered in solutions of benzene or cyclohexane with longer *n*-alkanes, which is probably due to size effects. Thus, in systems with benzene, $\Delta(\Delta G_m^*)$/J mol$^{-1}$ = −328 ($n$ = 6), 32 ($n$ =10), 828 ($n$ =16), and for cyclohexane mixtures, $\Delta(\Delta G_m^*)$ /J mol$^{-1}$ = −260 ($n$ = 6), 0 ($n$ =10), 796 ($n$ =16) (Table S3). It is remarkable that more negative values of $\Delta(\Delta G_m^*)$ are obtained for mixtures including short *n*-alkanes (table S3, Figure S2), which suggests that dispersive interactions are dominant [40].

*5.2.2 Correlation equations*

The application of the McAllister equation (two adjustable parameters) to the systems containing iodobenzene provides a mean relative standard deviation $\bar{\sigma}_r(v) = (1/N_S)\sum\sigma_r(v)$ = 0.002 ($N_S$ is the number of systems), which reveals an excellent agreement between experimental and calculated kinematic viscosities. Results are somewhat poorer when the Grunberg-Nissan equation is used (one adjustable parameter) since $\bar{\sigma}_r(\eta)$ (defined similarly to $\bar{\sigma}_r(v)$) is 0.013. The application of the equation (4) improves slightly results ($\bar{\sigma}_r(\eta)$ = 0.009). It seems that size effects are not very relevant on $\eta$ values in this type of systems. The same trend is hold for the remainder mixtures under consideration since both equations provide very similar results (Table S2). With regard to the Grunberg-Nissan equation, it must be remarked that the adjustable parameter is negative at any temperature for mixtures containing iodobenzene (Table 5), or 1-chloronaphthalene, or 1,2,4-trichlorobenzene or methyl benzoate (Table S2). Typically, negative values of $G_{12}$ have been ascribed to dispersive interactions are dominant [44,52]. In our previous studies on systems formed by one *n*-alkane and chlorobenzene, or bromobenzene, or 1,2,4-trichlorobenzene or 1-chloronaphthalene [10], we concluded that orientational effects are rather weak, and one can expect a similar behaviour in systems with iodobenzene. Finally, it should be noted that the $G_{12}$ parameter increases from negative to positive values in systems with benzene or cyclohexane (Table S2). For example, the systems with $n$ = 14,16, show positive values of the parameter. This underlines that $G_{12}$ is not only related to interactional effects, and that the size and shape of the mixture compounds and other structural effects may also contribute to its value.

*5.2.3 Bloomfield-Dewan's model*

Firstly, it must be noted that the free volume theory ($\alpha$ =1; $\beta$ =0) provides rather good results for most of the considered systems (Table 6, Figures 3-8). In fact, values of $\sigma_r(\eta)$ are lower than 6% for 20 mixtures, and only 4 solutions show deviations between experimental and calculated data larger than 10%, with a maximum deviation of 15%. This is acceptable since viscosities are determined from density measurements and no adjustable parameter is used. Poorer results are obtained for the mixtures: 1-chloronaphthalene + hexane ($\sigma_r(\eta)$ = 0.146), or + heptane ($\sigma_r(\eta)$ = 0.107), or for methyl benzoate + hexane ($\sigma_r(\eta)$ = 0.103), or for cyclohexane + hexane ($\sigma_r(\eta)$ = 0.099) (Table 6). Results show that, in this case, the model provides higher values of the theoretical viscosities than the experimental ones (see, e.g., Figures 4 and 6), which reveals that the contribution from free volume effects on viscosity is overestimated. Consequently, the absolute reaction rate contribution to $\ln \eta$ should be applied to improve results. It is important to note that this is attained calculating $\Delta G_m^{RES}$ by means of DISQUAC with interaction parameters available in the literature, and determined from phase equilibria and calorimetric data. Nevertheless, it must be remarked that, in this approach, values of $\Delta G_m^{RES}$ are overestimated since the combinatorial term, represented in the DISQUAC model, by the Flory-Huggins equation, is also overestimated, which is a shortcoming of the method applied.

We finish with a comment on the results obtained for the systems benzene or cyclohexane + tetradecane, or + hexadecane, characterized by rather similar free volume effects, which can be roughly evaluated by means of the difference between of the isobaric expansion coefficients ($\alpha_{pi}$) of the mixture components. Using values of $\alpha_{pi}$ from reference [74] and $\alpha_{pi}$ (benzene) = 1.213 $10^{-3}$ K$^{-1}$ [75], we have for benzene mixtures, ($\alpha_{p1}(C_6H_6) - \alpha_{p2}$)/$10^{-3}$ K$^{-1}$ = 0.327 ($n$ =14), 0.320 ($n$ =16) and for cyclohexane solutions, ($\alpha_{p1}(C_6H_{12}) - \alpha_{p2}$)/$10^{-3}$ K$^{-1}$ = 0.334 ($n$ =14), 0.337 ($n$ =16). The free volume model provides rather good results for the systems with benzene and slightly poorer for those containing cyclohexane (Table 6), a difference which may be ascribed to the different shape of these cyclic molecules.

## 6. Conclusions

Experimental values of $\nu$ and $\eta$ for iodobenzene + $n$-alkane mixtures at (288.15-308.15) K and atmospheric pressure are reported. These data and those from the literature for similar systems with 1-chloronaphthalene, or 1,2,4-trichlorobenzene, or methyl benzoate, or benzene, or cyclohexane have been examined using thermodynamic functions such as $U_{V_m}^E$, $TS_m^E$ or $V_m^E$. The observed loss of fluidization of mixtures involving iodobenzene, 1,2,4-trichlorobenzene, or 1-chloronaphthalene when $n$ increases can be ascribed to a decrease of the number of broken interactions between like molecules upon mixing. The disruption of CMO which exists in longer $n$-alkanes may explain the decreased negative values of $\Delta \eta$ for mixtures with benzene and $n =14,16$. The replacement, in this type of systems, of benzene by cyclohexane leads to increased positive $\Delta \eta$ values, probably due to the different shape of cyclohexane. Mixtures with iodobenzene, or 1,2,4-trichlorobenzene, or 1-chloronaphthalene or methyl benzoate and a short $n$-alkane are characterized by large structural effects and larger negative $\Delta \eta$ values. Results from the models reveal that dispersive interactions are dominant and that size effects are not relevant on $\eta$ values. The free volume model provides good results for most of the systems considered. Calculations taking into account simultaneously free volume effects and the absolute reaction rate model with $\Delta G_m^{RES}$ values determined using DISQUAC with interaction parameters obtained from low pressure phase equilibria data improves results for a number of solutions.

## Acknowledgements

This work was carried out under "Project PID2022-137104NA-I00", funded by MCIN/AEI/10.13039/501100011033/ and by FEDER Una manera de hacer Europa.

**TABLE 1**

Sample description.

| Chemical name | CAS Number | Source | Initial purity[a] |
|---|---|---|---|
| iodobenzene | 591-50-4 | Sigma-Aldrich | 0.999 |
| $n$-heptane | 142-82-5 | Fluka | 0.998 |
| $n$-decane | 124-18-5 | Sigma-Aldrich | 0.995 |
| $n$-dodecane | 112-40-3 | Sigma-Aldrich | 0.998 |
| $n$-tetradecane | 629-59-4 | Fluka | 0.995 |

[a] In mole fraction. Initial purity, measured by gas chromatography, certified by the supplier,

**TABLE 2**

Kinematic ($\nu$) and dynamic ($\eta$) viscosities of pure compounds at temperature $T$ and atmospheric pressure. Comparison of experimental (Exp.) results with literature (Lit.) values.[a]

| $T$/K | iodobenzene | | $n$-C$_7$ | | $n$-C$_{10}$ | | $n$-C$_{12}$ | | $n$-C$_{14}$ | |
|---|---|---|---|---|---|---|---|---|---|---|
| | \multicolumn{10}{c}{$\nu$ /cst} |
| | Exp. | Lit. | Exp. | Lit. | Exp. | Lit. | Exp. | Lit. | Exp. | Lit. |
| 288.15 | 0.983 | | 0.635 | 0.629[76] | 1.364 | 1.364[40] | 2.190 | 2.170[40] | 3.407 | 3.380[40] |
| | | | | | | 1.350[48] | | | | |
| 293.15 | 0.912 | | 0.612 | 0.600[77] | 1.263 | 1.254[58] | 2.001 | 1.982[58] | 3.058 | 3.020[79] |
| | | | | 0.604[78] | | 1.258[77] | | 1.974[77] | | 3.069[81] |
| | | | | | | 1.260[79] | | 1.960[79] | | 3.052[82] |
| | | | | | | | | 1.990[80] | | |
| 298.15 | 0.850 | | 0.584 | 0.574[76] | 1.180 | 1.169[40] | 1.837 | 1.809[40] | 2.768 | 2.753[81,88] |
| | | | | 0.572[77] | | 1.172[48] | | 1.818[58] | | 2.739[89] |
| | | | | 0.577[83] | | 1.168[58] | | 1.825[77] | | |
| | | | | 0.594[84] | | 1.186[75] | | | | |
| | | | | 0.582[85,86] | | 1.172[77] | | | | |
| | | | | | | 1.183[87] | | | | |
| 303.15 | 0.790 | | 0.543 | 0.549[76] | 1.103 | 1.096[58] | 1.694 | 1.682[58] | 2.516 | 2.480[79] |
| | | | | 0.551[83] | | 1.100[79] | | 1.670[79] | | 2.512[82] |
| | | | | | | | | 1.680[80] | | |
| 308.15 | 0.742 | | 0.527 | 0.532[76] | 1.035 | 1.034[40] | 1.569 | 1.557[40] | 2.297 | 2.271[40] |
| | | | | 0.522[77] | | | | 1.558[77] | | 2.277[91] |
| | | | | 0.528[83] | | | | | | |
| | | | | 0.521[90] | | | | | | |
| | \multicolumn{10}{c}{$\eta$ /mPa s} |
| | iodobenzene | | $n$-C$_7$ | | $n$-C$_{10}$ | | $n$-C$_{12}$ | | $n$-C$_{14}$ | |
| | Exp. | Lit. | Exp. | Lit. | Exp. | Lit. | Exp. | Lit. | Exp. | Lit. |
| 288.15 | 1.805 | 1.740[92] | 0.437 | 0.433[76] | 0.998 | 0.991[48] | 1.648 | 1.636[40] | 2.611 | 2.596[40] |
| | | | | | | 1.003[76] | | 1.635[95] | | 2.589[96] |
| 293.15 | 1.669 | 1.736[93] | 0.418 | 0.410[76,53] | 0.922 | 0.916[58] | 1.499 | 1.485[58] | 2.333 | 2.300[79] |
| | | | | 0.411[77] | | 0.918[77] | | 1.480[77] | | 2.289[96] |
| | | | | 0.413[94] | | 0.922[79] | | 1.470[79] | | 2.330[82] |

TABLE 2 (continued)

| T | | | | | | | | | | |
|---|---|---|---|---|---|---|---|---|---|---|
| | | | | | | | | | 1.490[80] | |
| | | | | | | | | | 1.496[95] | |
| | 1.548 | | 0.397 | 0.390[76] | 0.857 | 0.849[40] | 1.369 | 1.348[40] | 2.102 | 2.101[14] |
| 298.15 | | | | 0.389[77] | | 0.851[48,77] | | 1.356[58] | | 2.085[88] |
| | | | | 0.392[83] | | 0.848[58] | | 1.360[77] | | 2.091[60] |
| | | | | 0.404[84] | | 0.861[75] | | 1.364[95] | | 2.087[96] |
| | | | | 0.390[97] | | 0.859[87] | | 1.367[60] | | |
| | | | | 0.388[98] | | 0.859[99] | | | | |
| | | | | 0.393[16] | | | | | | |
| | | | | 0.396[86] | | | | | | |
| 303.15 | 1.434 | 1.417[92] | 0.367 | 0.371[76] | 0.797 | 0.792[58] | 1.252 | 1.248[58] | 1.902 | 1.895[14] |
| | | | | 0.373[83] | | 0.794[79] | | 1.240[79] | | 1.900[82] |
| | | | | 0.368[98] | | 0.800[101] | | 1.250[80] | | |
| | | | | 0.363[100] | | | | 1.254[95] | | |
| 308.15 | 1.341 | | 0.354 | 0.357[76] | 0.744 | 0.743[76] | 1.158 | 1.149[40] | 1.728 | 1.715[14] |
| | | | | 0.351[77] | | 0.741[77] | | 1.150[77] | | 1.709[40] |
| | | | | 0.354[83] | | | | 1.157[95] | | 1.717[96] |
| | | | | 0.350[90,98] | | | | | | |
| | | | | 0.352[97] | | | | | | |

[a] The uncertainties are: $u(T) = \pm 0.02$ K; $u(p) = \pm 1$ kPa. For viscosities, the relative combined expanded uncertainties (0.95 level of confidence) are $U_{rc}(v) = 0.020$ and $U_{rc}(\eta) = 0.022$

**TABLE 3**

Kinematic ($\nu$) and dynamic ($\eta$) viscosities of iodobenzene (1) + n-alkane (2) mixtures at temperature $T$ and atmospheric pressure. Values of $\Delta\eta$ (eq. 9) are also given[a].

| $x_1$ | $\nu$/cst | $\eta$/mPa s | $\Delta\eta$/mPa s |
|---|---|---|---|
| | iodobenzene (1) + heptane (2); $T$/K = 288.15 | | |
| 0.0504 | 0.626 | 0.459 | −0.046 |
| 0.0951 | 0.617 | 0.478 | −0.089 |
| 0.1454 | 0.614 | 0.504 | −0.131 |
| 0.1925 | 0.609 | 0.527 | −0.172 |
| 0.2441 | 0.609 | 0.558 | −0.211 |
| 0.2929 | 0.611 | 0.590 | −0.246 |
| 0.3940 | 0.622 | 0.667 | −0.307 |
| 0.4954 | 0.648 | 0.767 | −0.346 |
| 0.5916 | 0.676 | 0.876 | −0.368 |
| 0.6936 | 0.728 | 1.033 | −0.350 |
| 0.7452 | 0.751 | 1.116 | −0.338 |
| 0.7958 | 0.789 | 1.224 | −0.299 |
| 0.8450 | 0.819 | 1.326 | −0.264 |
| 0.8963 | 0.866 | 1.463 | −0.197 |
| 0.9480 | 0.917 | 1.615 | −0.115 |
| | iodobenzene (1) + heptane (2); $T$/K = 293.15 | | |
| 0.0504 | 0.604 | 0.440 | −0.041 |
| 0.0951 | 0.596 | 0.458 | −0.079 |
| 0.1454 | 0.592 | 0.483 | −0.117 |
| 0.1925 | 0.587 | 0.506 | −0.154 |
| 0.2441 | 0.587 | 0.535 | −0.189 |
| 0.2929 | 0.588 | 0.565 | −0.220 |
| 0.3940 | 0.598 | 0.638 | −0.274 |
| 0.4954 | 0.621 | 0.731 | −0.307 |
| 0.5916 | 0.645 | 0.832 | −0.327 |
| 0.6936 | 0.685 | 0.969 | −0.318 |
| 0.7452 | 0.711 | 1.051 | −0.300 |
| 0.7958 | 0.745 | 1.151 | −0.263 |
| 0.8450 | 0.771 | 1.242 | −0.235 |
| 0.8963 | 0.815 | 1.371 | −0.170 |
| 0.9480 | 0.855 | 1.500 | −0.105 |
| | iodobenzene (1) + heptane (2); $T$/K = 298.15 | | |

TABLE 3 (continued)

| | | | |
|---|---|---|---|
| 0.0504 | 0.576 | 0.417 | −0.038 |
| 0.0951 | 0.568 | 0.434 | −0.072 |
| 0.1454 | 0.564 | 0.458 | −0.107 |
| 0.1925 | 0.560 | 0.479 | −0.140 |
| 0.2441 | 0.559 | 0.507 | −0.172 |
| 0.2929 | 0.560 | 0.535 | −0.200 |
| 0.3940 | 0.570 | 0.605 | −0.247 |
| 0.4954 | 0.586 | 0.686 | −0.283 |
| 0.5916 | 0.610 | 0.783 | −0.298 |
| 0.6936 | 0.646 | 0.908 | −0.290 |
| 0.7452 | 0.670 | 0.987 | −0.271 |
| 0.7958 | 0.704 | 1.083 | −0.233 |
| 0.8450 | 0.729 | 1.168 | −0.205 |
| 0.8963 | 0.765 | 1.280 | −0.152 |
| 0.9480 | 0.803 | 1.402 | −0.089 |
| iodobenzene (1) + heptane (2); $T$/K = 303.15 | | | |
| 0.0504 | 0.535 | 0.385 | −0.035 |
| 0.0951 | 0.527 | 0.401 | −0.067 |
| 0.1454 | 0.523 | 0.422 | −0.099 |
| 0.1925 | 0.518 | 0.441 | −0.130 |
| 0.2441 | 0.518 | 0.467 | −0.159 |
| 0.2929 | 0.520 | 0.495 | −0.183 |
| 0.3940 | 0.527 | 0.556 | −0.229 |
| 0.4954 | 0.542 | 0.632 | −0.260 |
| 0.5916 | 0.565 | 0.721 | −0.272 |
| 0.6936 | 0.598 | 0.837 | −0.265 |
| 0.7452 | 0.620 | 0.908 | −0.248 |
| 0.7958 | 0.649 | 0.994 | −0.216 |
| 0.8450 | 0.674 | 1.075 | −0.187 |
| 0.8963 | 0.705 | 1.175 | −0.142 |
| 0.9480 | 0.742 | 1.290 | −0.081 |
| iodobenzene (1) + heptane (2); $T$/K = 308.15 | | | |
| 0.0504 | 0.519 | 0.371 | −0.032 |
| 0.0951 | 0.511 | 0.386 | −0.061 |
| 0.1454 | 0.507 | 0.407 | −0.090 |
| 0.1925 | 0.502 | 0.425 | −0.118 |
| 0.2441 | 0.501 | 0.449 | −0.145 |
| 0.2929 | 0.502 | 0.474 | −0.168 |
| 0.3940 | 0.511 | 0.536 | −0.206 |
| 0.4954 | 0.524 | 0.608 | −0.235 |
| 0.5916 | 0.544 | 0.690 | −0.247 |
| 0.6936 | 0.573 | 0.798 | −0.240 |
| 0.7452 | 0.594 | 0.866 | −0.223 |
| 0.7958 | 0.617 | 0.941 | −0.198 |
| 0.8450 | 0.640 | 1.017 | −0.170 |
| 0.8963 | 0.670 | 1.112 | −0.126 |
| 0.9480 | 0.702 | 1.215 | −0.074 |
| iodobenzene (1) + decane (2); $T/K = 288.15$ | | | |
| 0.0968 | 1.268 | 1.012 | −0.064 |
| 0.1484 | 1.224 | 1.022 | −0.096 |
| 0.1980 | 1.185 | 1.032 | −0.126 |
| 0.2594 | 1.143 | 1.050 | −0.158 |
| 0.2959 | 1.120 | 1.062 | −0.176 |
| 0.3454 | 1.091 | 1.080 | −0.198 |
| 0.3939 | 1.065 | 1.099 | −0.218 |
| 0.4946 | 1.020 | 1.151 | −0.247 |
| 0.5945 | 0.986 | 1.218 | −0.261 |
| 0.6969 | 0.961 | 1.307 | −0.255 |
| 0.7461 | 0.954 | 1.361 | −0.240 |
| 0.7960 | 0.949 | 1.421 | −0.221 |
| 0.8465 | 0.950 | 1.495 | −0.188 |
| 0.8974 | 0.952 | 1.575 | −0.149 |
| 0.9486 | 0.964 | 1.679 | −0.086 |
| iodobenzene (1) + decane (2); $T/K = 293.15$ | | | |
| 0.0968 | 1.180 | 0.938 | −0.057 |
| 0.1484 | 1.140 | 0.947 | −0.086 |
| 0.1980 | 1.103 | 0.957 | −0.113 |
| 0.2594 | 1.066 | 0.974 | −0.142 |
| 0.2959 | 1.044 | 0.985 | −0.158 |
| 0.3454 | 1.017 | 1.001 | −0.178 |
| 0.3939 | 0.994 | 1.020 | −0.195 |

TABLE 3 (continued)

| | | | |
|---|---|---|---|
| 0.4946 | 0.952 | 1.069 | −0.221 |
| 0.5945 | 0.920 | 1.132 | −0.233 |
| 0.6969 | 0.896 | 1.213 | −0.228 |
| 0.7461 | 0.889 | 1.262 | −0.216 |
| 0.7960 | 0.885 | 1.319 | −0.196 |
| 0.8465 | 0.884 | 1.385 | −0.168 |
| 0.8974 | 0.887 | 1.461 | −0.130 |
| 0.9486 | 0.895 | 1.553 | −0.076 |
| iodobenzene (1) + decane (2); $T$/K = 298.15 | | | |
| 0.0968 | 1.103 | 0.872 | −0.052 |
| 0.1484 | 1.066 | 0.882 | −0.078 |
| 0.1980 | 1.033 | 0.891 | −0.102 |
| 0.2594 | 0.999 | 0.908 | −0.127 |
| 0.2959 | 0.978 | 0.918 | −0.143 |
| 0.3454 | 0.954 | 0.934 | −0.161 |
| 0.3939 | 0.931 | 0.952 | −0.177 |
| 0.4946 | 0.892 | 0.997 | −0.200 |
| 0.5945 | 0.863 | 1.057 | −0.210 |
| 0.6969 | 0.841 | 1.134 | −0.203 |
| 0.7461 | 0.832 | 1.176 | −0.195 |
| 0.7960 | 0.830 | 1.232 | −0.173 |
| 0.8465 | 0.828 | 1.291 | −0.149 |
| 0.8974 | 0.828 | 1.358 | −0.117 |
| 0.9486 | 0.835 | 1.442 | −0.068 |
| iodobenzene (1) + decane (2); $T$/K = 303.15 | | | |
| 0.0968 | 1.033 | 0.813 | −0.046 |
| 0.1484 | 0.999 | 0.822 | −0.070 |
| 0.1980 | 0.969 | 0.832 | −0.092 |
| 0.2594 | 0.937 | 0.848 | −0.115 |
| 0.2959 | 0.919 | 0.858 | −0.128 |
| 0.3454 | 0.896 | 0.873 | −0.145 |
| 0.3939 | 0.874 | 0.889 | −0.160 |
| 0.4946 | 0.839 | 0.933 | −0.180 |
| 0.5945 | 0.811 | 0.989 | −0.188 |
| 0.6969 | 0.790 | 1.060 | −0.183 |
| 0.7461 | 0.783 | 1.101 | −0.173 |
| 0.7960 | 0.780 | 1.153 | −0.153 |
| 0.8465 | 0.779 | 1.210 | −0.129 |
| 0.8974 | 0.776 | 1.268 | −0.102 |
| 0.9486 | 0.783 | 1.347 | −0.057 |
| iodobenzene (1) + decane (2); $T$/K = 308.15 | | | |
| 0.0968 | 0.971 | 0.760 | −0.042 |
| 0.1484 | 0.940 | 0.769 | −0.064 |
| 0.1980 | 0.911 | 0.778 | −0.084 |
| 0.2594 | 0.882 | 0.794 | −0.105 |
| 0.2959 | 0.864 | 0.802 | −0.119 |
| 0.3454 | 0.843 | 0.818 | −0.133 |
| 0.3939 | 0.823 | 0.833 | −0.146 |
| 0.4946 | 0.791 | 0.875 | −0.164 |
| 0.5945 | 0.763 | 0.925 | −0.174 |
| 0.6969 | 0.744 | 0.994 | −0.167 |
| 0.7461 | 0.737 | 1.033 | −0.157 |
| 0.7960 | 0.735 | 1.081 | −0.139 |
| 0.8465 | 0.738 | 1.141 | −0.109 |
| 0.8974 | 0.735 | 1.196 | −0.085 |
| 0.9486 | 0.735 | 1.259 | −0.052 |
| iodobenzene (1) + dodecane (2); $T$/K = 288.15 | | | |
| 0.0988 | 1.988 | 1.609 | −0.055 |
| 0.1496 | 1.895 | 1.594 | −0.079 |
| 0.1964 | 1.810 | 1.576 | −0.104 |
| 0.2466 | 1.728 | 1.562 | −0.126 |
| 0.2971 | 1.648 | 1.547 | −0.149 |
| 0.3978 | 1.505 | 1.529 | −0.183 |
| 0.4966 | 1.381 | 1.522 | −0.205 |
| 0.5964 | 1.274 | 1.534 | −0.209 |
| 0.5978 | 1.271 | 1.533 | −0.210 |
| 0.6974 | 1.175 | 1.559 | −0.201 |
| 0.7480 | 1.132 | 1.579 | −0.189 |
| 0.7977 | 1.096 | 1.609 | −0.167 |
| 0.8486 | 1.061 | 1.644 | −0.139 |
| 0.8982 | 1.030 | 1.685 | −0.106 |
| 0.9487 | 1.002 | 1.733 | −0.066 |

TABLE 3 (continued)

iodobenzene (1) + dodecane (2); $T/K = 293.15$

| | | | |
|---|---|---|---|
| 0.0988 | 1.821 | 1.467 | −0.048 |
| 0.1496 | 1.738 | 1.455 | −0.069 |
| 0.1964 | 1.663 | 1.441 | −0.091 |
| 0.2466 | 1.590 | 1.430 | −0.110 |
| 0.2971 | 1.519 | 1.419 | −0.130 |
| 0.3989 | 1.390 | 1.407 | −0.159 |
| 0.4966 | 1.277 | 1.402 | −0.181 |
| 0.5964 | 1.179 | 1.414 | −0.186 |
| 0.6974 | 1.091 | 1.441 | −0.175 |
| 0.7480 | 1.050 | 1.458 | −0.167 |
| 0.7977 | 1.016 | 1.485 | −0.148 |
| 0.8475 | 0.982 | 1.513 | −0.128 |
| 0.8974 | 0.953 | 1.552 | −0.098 |
| 0.9487 | 0.928 | 1.599 | −0.060 |

iodobenzene (1) + dodecane (2); $T/K = 298.15$

| | | | |
|---|---|---|---|
| 0.0988 | 1.676 | 1.344 | −0.042 |
| 0.1496 | 1.605 | 1.337 | −0.059 |
| 0.1964 | 1.534 | 1.323 | −0.080 |
| 0.2466 | 1.471 | 1.317 | −0.096 |
| 0.2971 | 1.404 | 1.305 | −0.116 |
| 0.3989 | 1.293 | 1.302 | −0.137 |
| 0.4966 | 1.190 | 1.299 | −0.157 |
| 0.5978 | 1.101 | 1.316 | −0.158 |
| 0.6956 | 1.020 | 1.338 | −0.153 |
| 0.7480 | 0.981 | 1.356 | −0.144 |
| 0.7977 | 0.947 | 1.379 | −0.131 |
| 0.8475 | 0.916 | 1.405 | −0.113 |
| 0.8974 | 0.889 | 1.440 | −0.087 |
| 0.9487 | 0.866 | 1.485 | −0.051 |

iodobenzene (1) + dodecane (2); $T/K = 303.15$

| | | | |
|---|---|---|---|
| 0.0988 | 1.550 | 1.237 | −0.033 |
| 0.1496 | 1.485 | 1.231 | −0.049 |
| 0.1964 | 1.422 | 1.221 | −0.067 |
| 0.2466 | 1.364 | 1.215 | −0.082 |
| 0.2971 | 1.304 | 1.207 | −0.100 |

TABLE 3 (continued)

| | | | |
|---|---|---|---|
| 0.3989 | 1.200 | 1.203 | −0.123 |
| 0.4966 | 1.105 | 1.201 | −0.142 |
| 0.5964 | 1.022 | 1.214 | −0.148 |
| 0.6956 | 0.950 | 1.241 | −0.140 |
| 0.7480 | 0.916 | 1.261 | −0.129 |
| 0.7977 | 0.887 | 1.286 | −0.114 |
| 0.8475 | 0.858 | 1.312 | −0.097 |
| 0.8982 | 0.835 | 1.349 | −0.070 |
| 0.9487 | 0.811 | 1.385 | −0.042 |
| iodobenzene (1) + dodecane (2); $T/K = 308.15$ | | | |
| 0.0988 | 1.439 | 1.142 | −0.034 |
| 0.1496 | 1.380 | 1.138 | −0.047 |
| 0.1964 | 1.323 | 1.130 | −0.064 |
| 0.2466 | 1.270 | 1.126 | −0.077 |
| 0.2971 | 1.216 | 1.121 | −0.092 |
| 0.3989 | 1.122 | 1.119 | −0.112 |
| 0.4966 | 1.034 | 1.119 | −0.131 |
| 0.5964 | 0.957 | 1.131 | −0.136 |
| 0.6956 | 0.892 | 1.160 | −0.126 |
| 0.7480 | 0.859 | 1.178 | −0.117 |
| 0.7977 | 0.834 | 1.203 | −0.102 |
| 0.8475 | 0.816 | 1.241 | −0.072 |
| 0.8982 | 0.785 | 1.262 | −0.061 |
| 0.9487 | 0.761 | 1.295 | −0.037 |
| iodobenzene (1) + tetradecane (2); $T/K = 288.15$ | | | |
| 0.1007 | 3.023 | 2.473 | −0.056 |
| 0.1478 | 2.857 | 2.411 | −0.081 |
| 0.1966 | 2.692 | 2.345 | −0.107 |
| 0.2479 | 2.531 | 2.280 | −0.131 |
| 0.2885 | 2.407 | 2.228 | −0.151 |
| 0.3485 | 2.236 | 2.156 | −0.174 |
| 0.3972 | 2.105 | 2.101 | −0.190 |
| 0.4991 | 1.851 | 1.999 | −0.209 |
| 0.5981 | 1.634 | 1.922 | −0.207 |
| 0.6981 | 1.436 | 1.860 | −0.188 |
| 0.7483 | 1.345 | 1.835 | −0.173 |

TABLE 3 (continued)

| | | | |
|---|---|---|---|
| 0.7982 | 1.261 | 1.816 | −0.151 |
| 0.8492 | 1.179 | 1.799 | −0.127 |
| 0.8986 | 1.111 | 1.796 | −0.091 |
| 0.9490 | 1.040 | 1.786 | −0.060 |
| iodobenzene (1) + tetradecane (2); $T$/K = 293.15 | | | |
| 0.1007 | 2.726 | 2.220 | −0.046 |
| 0.1478 | 2.582 | 2.169 | −0.066 |
| 0.1966 | 2.438 | 2.114 | −0.088 |
| 0.2479 | 2.297 | 2.059 | −0.109 |
| 0.2885 | 2.188 | 2.016 | −0.126 |
| 0.3485 | 2.038 | 1.956 | −0.146 |
| 0.3972 | 1.922 | 1.910 | −0.160 |
| 0.4991 | 1.697 | 1.825 | −0.177 |
| 0.5981 | 1.501 | 1.758 | −0.178 |
| 0.6981 | 1.323 | 1.706 | −0.164 |
| 0.7483 | 1.241 | 1.687 | −0.150 |
| 0.7982 | 1.165 | 1.671 | −0.132 |
| 0.8492 | 1.091 | 1.658 | −0.112 |
| 0.8986 | 1.028 | 1.655 | −0.082 |
| 0.9490 | 0.965 | 1.652 | −0.052 |
| iodobenzene (1) + tetradecane (2); $T$/K = 298.15 | | | |
| 0.1007 | 2.480 | 2.011 | −0.035 |
| 0.1478 | 2.351 | 1.966 | −0.054 |
| 0.1966 | 2.224 | 1.920 | −0.073 |
| 0.2479 | 2.102 | 1.876 | −0.088 |
| 0.2885 | 2.004 | 1.837 | −0.105 |
| 0.3485 | 1.870 | 1.787 | −0.122 |
| 0.3972 | 1.767 | 1.748 | −0.134 |
| 0.4991 | 1.566 | 1.676 | −0.150 |
| 0.5981 | 1.391 | 1.623 | −0.148 |
| 0.6981 | 1.227 | 1.576 | −0.140 |
| 0.7483 | 1.152 | 1.559 | −0.129 |
| 0.7982 | 1.083 | 1.546 | −0.114 |
| 0.8492 | 1.017 | 1.539 | −0.093 |
| 0.8986 | 0.958 | 1.535 | −0.070 |
| 0.9490 | 0.901 | 1.534 | −0.042 |

TABLE 3 (continued)

iodobenzene (1) + tetradecane (2); $T$/K = 303.15

| $x_1$ | $\eta$ | | $\Delta\eta$ |
|---|---|---|---|
| 0.1007 | 2.262 | 1.825 | −0.030 |
| 0.1478 | 2.150 | 1.789 | −0.044 |
| 0.1966 | 2.037 | 1.750 | −0.060 |
| 0.2479 | 1.928 | 1.713 | −0.074 |
| 0.2885 | 1.841 | 1.680 | −0.087 |
| 0.3485 | 1.721 | 1.636 | −0.104 |
| 0.3972 | 1.629 | 1.604 | −0.125 |
| 0.4991 | 1.449 | 1.544 | −0.128 |
| 0.5981 | 1.289 | 1.496 | −0.118 |
| 0.6981 | 1.141 | 1.459 | −0.108 |
| 0.7483 | 1.073 | 1.446 | −0.094 |
| 0.7982 | 1.010 | 1.437 | −0.076 |
| 0.8492 | 0.950 | 1.431 | −0.054 |
| 0.8986 | 0.896 | 1.430 | −0.037 |
| 0.9490 | 0.839 | 1.423 | −0.018 |

iodobenzene (1) + tetradecane (2); $T$/K = 308.15

| | | | |
|---|---|---|---|
| 0.1007 | 2.072 | 1.664 | −0.025 |
| 0.1478 | 1.972 | 1.634 | −0.037 |
| 0.1966 | 1.873 | 1.602 | −0.050 |
| 0.2479 | 1.775 | 1.570 | −0.062 |
| 0.2885 | 1.699 | 1.544 | −0.073 |
| 0.3485 | 1.591 | 1.507 | −0.087 |
| 0.3972 | 1.507 | 1.477 | −0.098 |
| 0.4991 | 1.344 | 1.426 | −0.109 |
| 0.5981 | 1.200 | 1.387 | −0.111 |
| 0.6981 | 1.065 | 1.355 | −0.103 |
| 0.7483 | 1.003 | 1.345 | −0.094 |
| 0.7982 | 0.943 | 1.336 | −0.084 |
| 0.8492 | 0.889 | 1.333 | −0.067 |
| 0.8986 | 0.835 | 1.327 | −0.055 |
| 0.9490 | 0.787 | 1.330 | −0.032 |

[a] the uncertainties, $u$, are: $u(T) = \pm 0.02$ K; $u(p) = \pm 1$ kPa ; $u(x_1) = \pm 0.0010$. The relative combined expanded uncertainties (0.95 level of confidence) for $\eta$ and $\Delta\eta$ are, respectively, $U_{rc}(\eta) = 0.022$ and $U_{rc}(\Delta\eta) = 0.050$.

**TABLE 4**

Coefficients $A_i$ and standard deviations, $\sigma(\Delta F)$ (eq. 11) for representation of the $Q^a$ property at temperature $T$ and atmospheric pressure for iodobenzene (1) + $n$-alkane (2) systems by eq. 10 (between parentheses, standard deviations of the coefficients are given).

| Property ($Q$) | $T$/K | $A_0$ | $A_1$ | $A_2$ | $A_3$ | $\sigma(Q)$ |
|---|---|---|---|---|---|---|
| \multicolumn{7}{c}{iodobenzene (1) + heptane (2)} |
| $\Delta \eta$ /mPa s | 288.15 | −1.392 | −0.56 | −0.29 | −0.22 | 0.003 |
| | | (±0.006) | (±0.02) | (±0.02) | (±0.05) | |
| | 293.15 | −1.243 | −0.57 | −0.24 | | 0.004 |
| | | (±0.007) | (±0.01) | (±0.03) | | |
| | 298.15 | −1.137 | −0.49 | −0.16 | | 0.003 |
| | | (±0.007) | (±0.01) | (±0.02) | | |
| | 303.15 | −1.043 | −0.447 | −0.16 | | 0.002 |
| | | (±0.005) | (±0.08) | (±0.02) | | |
| | 308.15 | −0.945 | −0401 | −0.15 | | 0.001 |
| | | (±0.004) | (±0.006) | (±0.01) | | |
| $\Delta(\Delta G_m^*)$/J mol$^{-1}$ | 298.15 | −1235 | −167 | | | 5 |
| | | (±8) | (±19) | | | |
| \multicolumn{7}{c}{iodobenzene (1) + decane (2)} |
| $\Delta \eta$ /mPa s | 288.15 | −0.988 | −0.41 | −0.26 | −0.21 | 0.002 |
| | | (±0.004) | (±0.01) | (±0.02) | (±0.03) | |
| | 293.15 | −0.886 | −0.368 | −0.223 | −0.16 | 0.001 |
| | | (±0.002) | (±0.007) | (±0.009) | (±0.02) | |
| | 298.15 | −0.800 | −0.32 | −0.185 | −0.15 | 0.002 |
| | | (±0.004) | (±0.01) | (±0.01) | (±0.03) | |
| | 303.15 | −0.723 | −0.319 | −0.13 | | 0.002 |
| | | (±0.004) | (±0.007) | (±0.02) | | |
| | 308.15 | −0.666 | −0.273 | −0.06 | | 0.003 |
| | | (±0.005) | (±0.009) | (±0.02) | | |
| $\Delta(\Delta G_m^*)$/J mol$^{-1}$ | 298.15 | −992 | −298 | −154 | | 2 |
| | | (±6) | (±10) | (±24) | | |
| \multicolumn{7}{c}{iodobenzene (1) + dodecane (2)} |
| $\Delta \eta$ /mPa s | 288.15 | −0.816 | −0.27 | −0.10 | −0.13 | 0.002 |
| | | (±0.005) | (±0.02) | (±0.02) | (±0.04) | |
| | 293.15 | −0.714 | −0.249 | −0.12 | −0.16 | 0.002 |
| | | (±0.005) | (±0.02) | (±0.02) | (±0.04) | |

TABLE 4 (continued)

| | T/K | $a_0$ | $a_1$ | $a_2$ | $a_3$ | $\sigma(Q)^a$ |
|---|---|---|---|---|---|---|
| | 298.15 | −0.617 (±0.004) | −0.19 (±0.01) | −0.12 (±0.02) | −0.18 (±0.03) | 0.002 |
| | 303.15 | −0.568 (±0.002) | −0.245 (±0.005) | | | 0.001 |
| | 308.15 | −0.516 (±0.006) | −0.19 (±0.01) | | | 0.004 |
| $\Delta(\Delta G_m^*)$/J mol$^{-1}$ | 298.15 | −479 (±7) | −245 (±11) | −162 (±28) | | 2 |
| iodobenzene (1) + tetradecane (2) | | | | | | |
| $\Delta\eta$ /mPa s | 288.15 | −0.826 (±0.006) | −0.16 (±0.02) | | −0.18 (±0.05) | 0.003 |
| | 293.15 | −0.702 (±0.004) | −0.17 (±0.02) | | −0.17 (±0.04) | 0.002 |
| | 298.15 | −0.589 (±0.004) | −0.206 (±0.009) | | | 0.003 |
| | 303.15 | −0.512 (±0.008) | −0.07 (±0.01) | 0.24 (±0.03) | | 0.004 |
| | 308.15 | −0.428 (±0.004) | −0.170 (±0.008) | | | 0.002 |
| $\Delta(\Delta G_m^*)$/J mol$^{-1}$ | 298.15 | 198 (±5) | −120 (±9) | −147 (±22) | | 3 |

$^a Q = \Delta\eta$ or $\Delta G_m^*$

**TABLE 5**

Results provided by the application of the Grunberg-Nissan model (eq. 3; adjustable parameter: ($G_{12}$)) and McAllister (eq. 2; adjustable parameters: $Z_{12}$ and $Z_{21}$) to iodobenzene (1) + n-alkane (2) mixtures at temperature $T$ and atmospheric pressure.

| Equation | | $T = 288.15$ K | | $T = 293.15$ K | | $T = 298.15$ K | | $T = 303.15$ K | | $T = 308.15$ K | |
|---|---|---|---|---|---|---|---|---|---|---|---|
| | | Param.[a] | $\sigma_r$[b] | Param.[a] | $\sigma_r$[b] | Param.[a] | $\sigma_r$[b] | Param.[a] | $\sigma_r$[b] | Param.[a] | $\sigma_r$[b] |
| Iodobenzene (1) + heptane (2) | | | | | | | | | | | |
| Grunberg-Nissan | $G_{12}$ | −0.593 | 0.006 | −0.554 | 0.006 | −0.526 | 0.004 | −0.531 | 0.003 | −0.498 | 0.003 |
| McAllister | $Z_{12}$ | 0.654 | 0.002 | 0.623 | 0.002 | 0.593 | 0.002 | 0.549 | 0.001 | 0.529 | 0.001 |
| | $Z_{21}$ | 0.591 | | 0.572 | | 0.543 | | 0.501 | | 0.486 | |
| Iodobenzene (1) + decane (2) | | | | | | | | | | | |
| Grunberg-Nissan | $G_{12}$ | −0.634 | 0.018 | −0.608 | 0.015 | −0.583 | 0.013 | −0.552 | 0.010 | −0.530 | 0.008 |
| McAllister | $Z_{12}$ | 0.882 | 0.003 | 0.828 | 0.003 | 0.781 | 0.002 | 0.741 | 0.002 | 0.704 | 0.001 |
| | $Z_{21}$ | 1.082 | | 1.009 | | 0.946 | | 0.887 | | 0.833 | |
| Iodobenzene (1) + dodecane (2) | | | | | | | | | | | |
| Grunberg-Nissan | $G_{12}$ | −0.510 | 0.020 | −0.487 | 0.002 | −0.456 | 0.017 | −0.435 | 0.015 | −0.428 | 0.012 |
| McAllister | $Z_{12}$ | 1.140 | 0.003 | 1.058 | 0.003 | 0.996 | 0.004 | 0.930 | 0.002 | 0.880 | 0.002 |
| | $Z_{21}$ | 1.591 | | 1.471 | | 1.363 | | 1.266 | | 1.178 | |
| Iodobenzene (1) + tetradecane (2) | | | | | | | | | | | |
| Grunberg-Nissan | $G_{12}$ | −0.310 | 0.030 | −0.298 | 0.020 | −0.278 | 0.200 | −0.258 | 0.017 | −0.248 | 0.016 |
| McAllister | $Z_{12}$ | 1.481 | 0.003 | 1.365 | 0.003 | 1.271 | 0.003 | 1.189 | 0.002 | 1.110 | 0.002 |
| | $Z_{21}$ | 2.314 | | 2.112 | | 1.938 | | 1.781 | | 1.647 | |

[a]adjustable parameter; [b]standard relative deviation (eq. 12)

**TABLE 6**

Dynamic viscosities ($\eta$), deviations in absolute viscosity ($\Delta\eta$, eq. 9) and excess molar volumes, $V_m^E$, for solute (1) + $n$-alkane (2) mixtures at 298.15 K, atmospheric pressure and equimolar composition. Results from the application of the Bloomfield-Dewan´s model (eq. 5) to these systems are also included.

| $n$-alkane | $N^a$ | $\eta$ / mPa s | $\Delta\eta$ / mPa s | $\alpha^b$ | $\beta^c$ | $\sigma_r(\eta)^d$ | Ref. $\eta$ | $V_m^E$ / cm$^3$ mol$^{-1}$ | Ref. $V_m^E$ |
|---|---|---|---|---|---|---|---|---|---|
| \multicolumn{10}{c}{iodobenzene (1) + $n$-C$_n$ (2)} |
| $n$-C$_7$ | 15 | 0.688 | −0.284 | 1 | 0 | 0.015 | This work | −0.467 | 12 |
| $n$-C$_{10}$ | 15 | 1.002 | −0.200 | 1 | 0 | 0.016 | This work | 0.056 | 12 |
| $n$-C$_{12}$ | 14 | 1.304 | −0.154 | 1 | 0 | 0.021 | This work | 0.235 | 12 |
| $n$-C$_{14}$ | 15 | 1.677 | −0.147 | 1 | 0 | 0.046 | This work | 0.372 | 12 |
| \multicolumn{10}{c}{1-chloronaphthalene (1) + $n$-C$_n$ (2)} |
| $n$-C$_6$ | 9 | 0.802 | −0.871 | 1 | 0 | 0.146 | 16 | −1.560 | 102 |
|  |  |  |  | 1 | 0.8 | 0.040 |  |  |  |
| $n$-C$_7$ | 9 | 0.937 | −0.796 | 1 | 0 | 0.107 | 16 | −1.263 | 102 |
|  |  |  |  | 1 | 0.6 | 0.027 |  |  |  |
| $n$-C$_8$ | 9 | 1.056 | −0.738 | 1 | 0 | 0.077 | 16 | −1.052 | 102 |
|  |  |  |  | 1 | 0.4 | 0.028 |  |  |  |
| $n$-C$_{10}$ | 9 | 1.357 | −0.607 | 1 | 0 | 0.055 | 16 | −0.749 | 16 |
|  |  |  |  | 1 | 0.25 | 0.024 |  |  |  |
| $n$-C$_{12}$ | 9 | 1.658 | −0.505 | 1 | 0 | 0.038 | 16 | −0.616 | 102 |
|  |  |  |  | 1 | 0.15 | 0.025 |  |  |  |
| \multicolumn{10}{c}{1,2,4-trichlorobenzene (1) + $n$-C$_n$ (2)} |
| $n$-C$_{10}$ | 9 | 1.204$^e$ | −0.284$^e$ | 1 | 0 | 0.018 | 15 | −0.180 | 15 |
|  |  |  |  | 1 | 0.1 | 0.005 |  |  |  |
| $n$-C$_{14}$ | 9 | 2.085$^e$ | −0.106$^e$ | 1 | 0 | 0.017 | 15 | 0.087 | 15 |
| \multicolumn{10}{c}{methyl benzoate (1) + $n$-C$_n$ (2)} |
| $n$-C$_6$ | 12 | 0.654 | −0.422 | 1 | 0 | 0.103 | 40 | −0.598 | 40 |
|  |  |  |  | 1 | 0.37 | 0.010 |  |  |  |
| $n$-C$_8$ | 17 | 0.802 | −0.369 | 1 | 0 | 0.083 | 40 | −0.071 | 40 |
|  |  |  |  | 1 | 0.25 | 0.007 |  |  |  |
| $n$-C$_{10}$ | 18 | 1.018 | −0.318 | 1 | 0 | 0.074 | 40 | 0.239 | 40 |

TABLE 6 (continued)

| | | | | | | | | | |
|---|---|---|---|---|---|---|---|---|---|
| | | | | | 1 | 0.2 | 0.005 | | |
| $n$-C$_{12}$ | 16 | 1.313 | −0.272 | 1 | 0 | 0.061 | 40 | 0.431 | 40 |
| | | | | 1 | 0.15 | 0.015 | | | |
| $n$-C$_{14}$ | 18 | 1.696 | −0.231 | 1 | 0 | 0.044 | 40 | 0.579 | 40 |
| | | | | 1 | 0.1 | 0.005 | | | |
| Cyclohexane (1) + $n$-C$_n$ (2) | | | | | | | | | |
| $n$-C$_6$ | 12 | 0.471 | −0.137 | 1 | 0 | 0.099 | 60 | 0.126 | 103 |
| | | | | 1 | 1 | 0.065 | | | |
| $n$-C$_7$ | 12 | 0.518 | −0.137 | 1 | 0 | 0.090 | 60 | 0.295 | 103 |
| | | | | 1 | 1 | 0.068 | | | |
| $n$-C$_8$ | 12 | 0.608 | −0.104 | 1 | 0 | 0.073 | 60 | 0.390 | 103 |
| | | | | 1 | 1 | 0.054 | | | |
| $n$-C$_{10}$ | 12 | 0.8298 | −0.047 | 1 | 0 | 0.044 | 60 | 0.494 | 103 |
| | | | | 1 | 1 | 0.025 | | | |
| $n$-C$_{12}$ | 12 | 1.134 | −0.003 | 1 | 0 | 0.028 | 60 | 0.556 | 104 |
| | | | | 1 | 0.7 | 0.013 | | | |
| $n$-C$_{14}$ | 12 | 1.526 | 0.028 | 1 | 0 | 0.039 | 60 | 0.591 | 105 |
| $n$-C$_{16}$ | 12 | 2.023 | 0.034 | 1 | 0 | 0.048 | 60 | 0.632 | 105 |
| benzene + $n$-C$_n$ | | | | | | | | | |
| $n$-C$_6$ | 10 | 0.372 | −0.088 | 1 | 0 | 0.053 | 59 | 0.403 | 106 |
| | | | | 1 | 0.3 | 0.013 | | | |
| $n$-C$_8$ | 9 | 0.490 | −0.075 | 1 | 0 | 0.056 | 59 | 0.700 | 106 |
| | | | | 1 | 0.3 | 0.015 | | | |
| $n$-C$_{10}$ | 10 | 0.665 | −0.060 | 1 | 0 | 0.032 | 59 | 0.846 | 107 |
| | | | | 1 | 0.25 | 0.009 | | | |
| $n$-C$_{12}$ | 10 | 0.898 | −0.093 | 1 | 0 | 0.033 | 59 | 0.919 | 73 |
| | | | | 1 | 0.5 | 0.008 | | | |
| $n$-C$_{14}$ | 9 | 1.184 | −0.126 | 1 | 0 | 0.018 | 59 | 1.015 | 59 |
| $n$-C$_{16}$ | 10 | 1.600 | −0.244 | 1 | 0 | 0.022 | 59 | 1.023 | 73 |

[a] number of data point; [b] weighing factor of $\ln \eta_{fv}$ to $\ln \eta$; [c] weighing factor of $\ln \eta_{ar}$ to $\ln \eta$; [d] equation (12); [e] values at 293.15 K

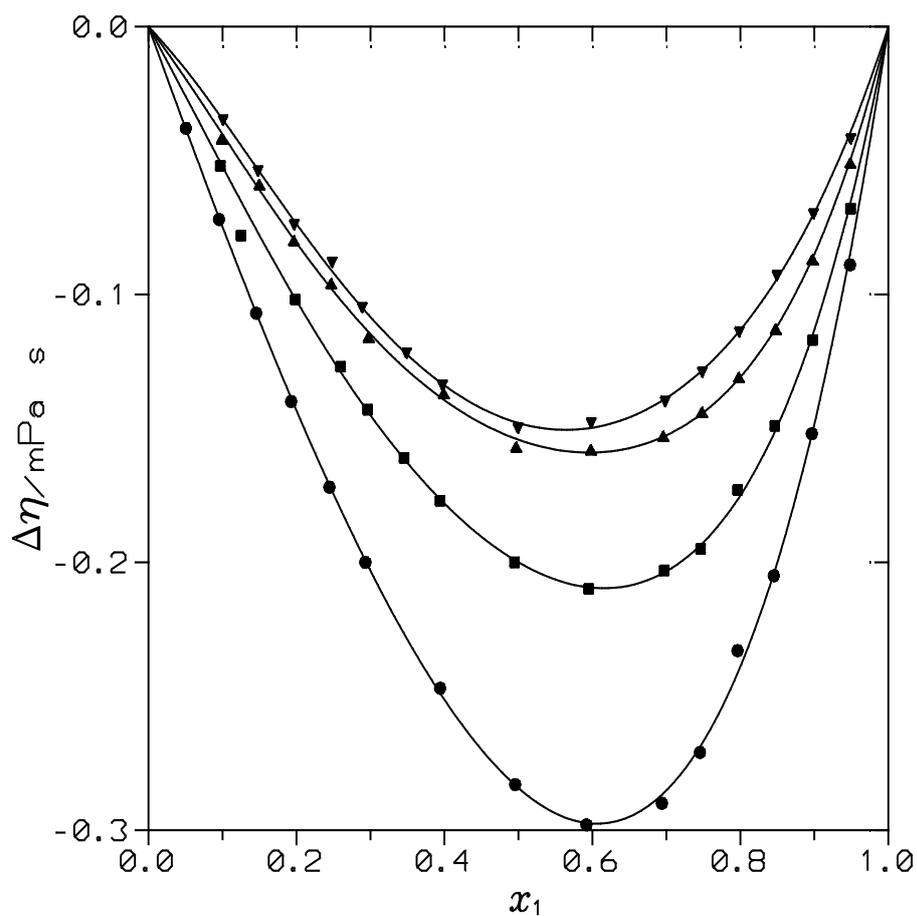

**Figure 1.** $\Delta \eta$ of iodobenzene (1) + *n*-alkane (2) mixtures at 298.15 K and atmospheric pressure. Points, experimental results (this work): (●), heptane; (■), decane, (▲); dodecane, (▼), tetradecane. Solid lines, calculations using coefficients listed in Table 4.

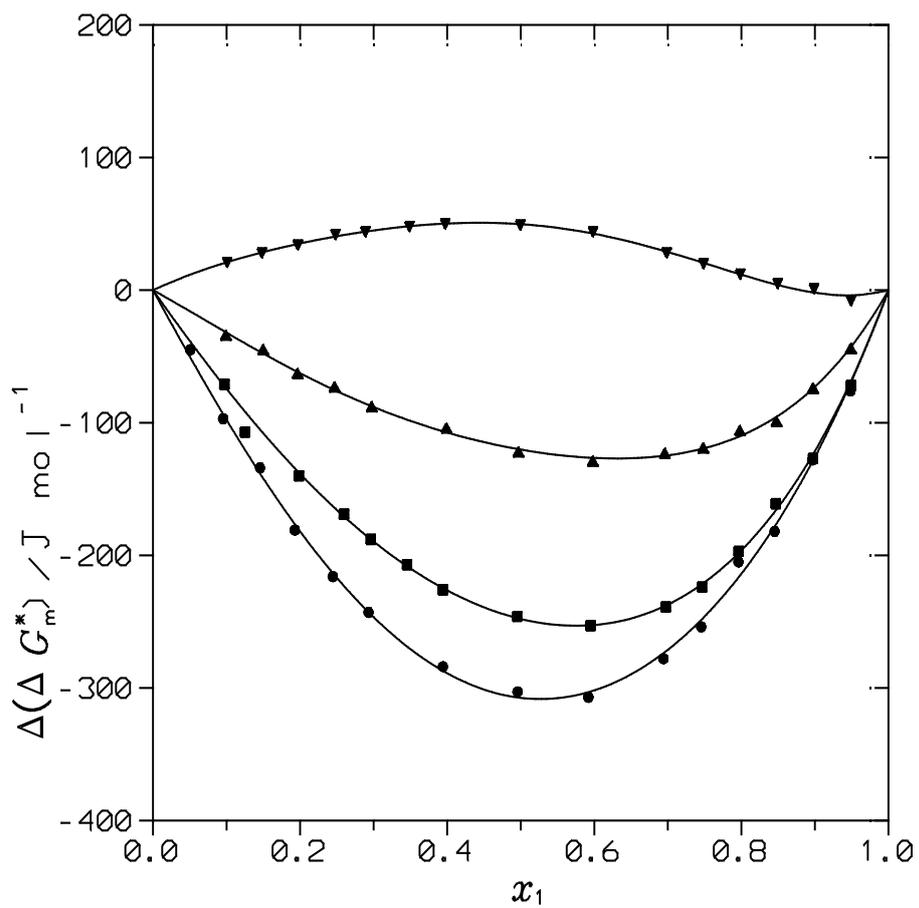

**Figure 2.** $\Delta(\Delta G_m^*)$ of iodobenzene (1) + $n$-alkane (2) mixtures at 298.15 K and atmospheric pressure. Points, experimental results (this work): (●), heptane; (■), decane; (▲), dodecane; (▼), tetradecane. Solid lines, calculations using coefficients listed in Table 4.

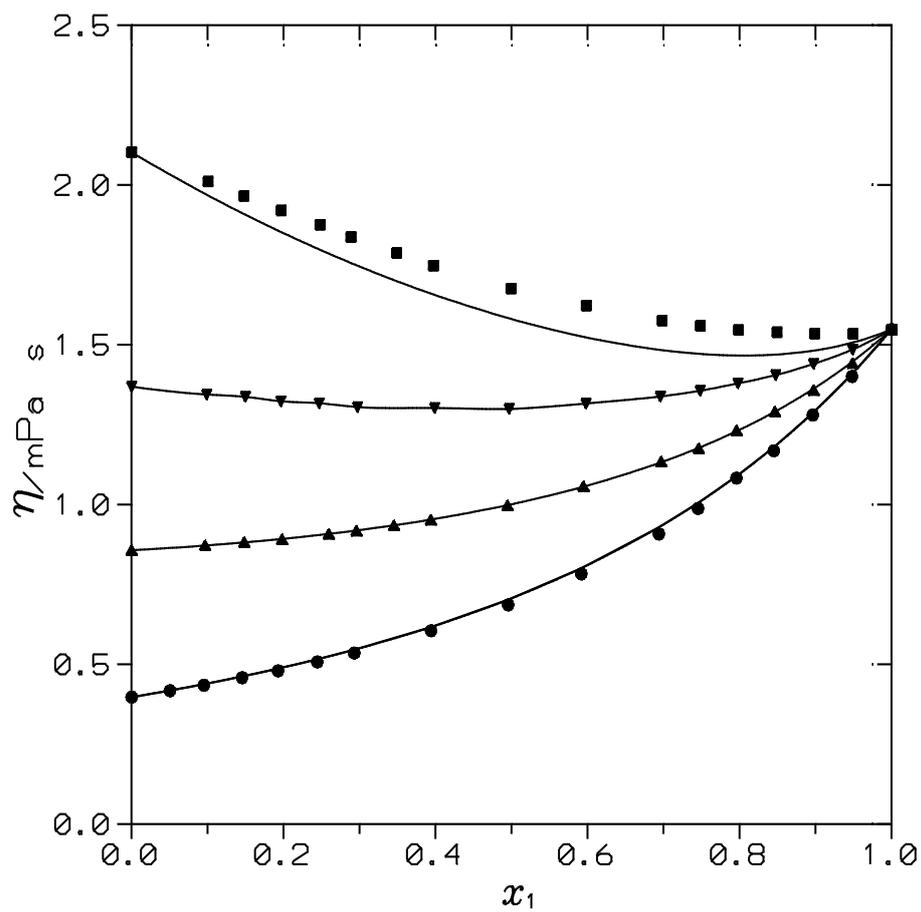

**Figure 3.**  Dynamic viscosities of iodobenzene (1) + *n*-alkane (2) mixtures at 298.15 K. Points, experimental results (this work): (●), heptane; (▲), decane; (▼), dodecane; (■), tetradecane. Solid lines, results from the application of the Bloomfield-Dewan´s model (equation 5) with ($\alpha = 1$, $\beta = 0$)

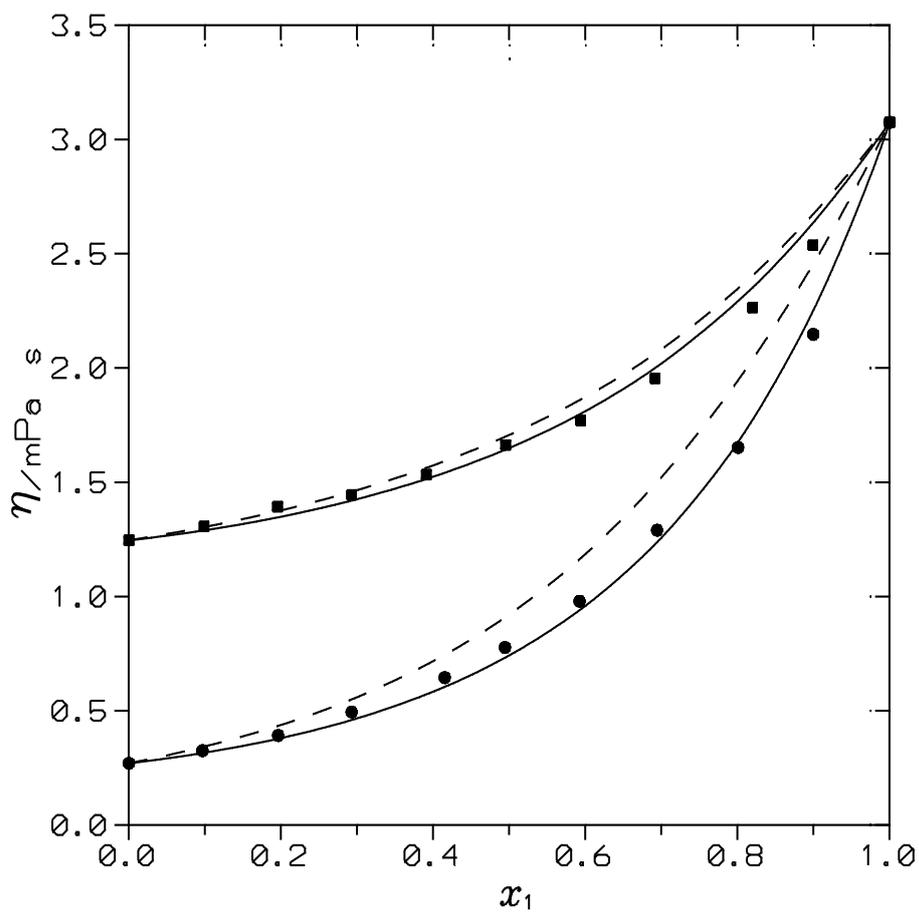

**Figure 4.** Dynamic viscosities of 1-chloronaphthalene (1) + $n$-alkane (2) mixtures at 298.15 K and atmospheric pressure. Points, experimental results [16]: (●), hexane; (■), dodecane. Dashed lines, results from the application of the Bloomfield-Dewan´s model (equation 5) with ($\alpha =1$, $\beta = 0$); solid lines, results using ($\alpha =1$, $\beta = 0.8$) for the mixture with hexane and ($\alpha =1$, $\beta = 0.15$) for the solution with dodecane.

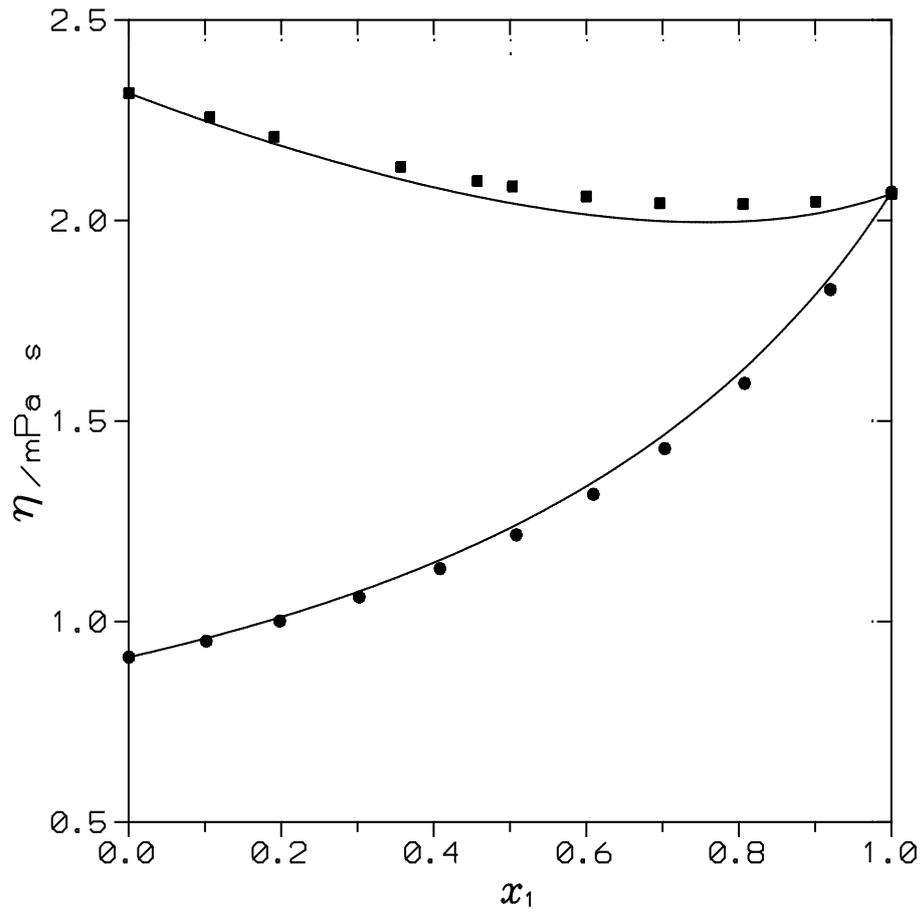

**Figure 5.** Dynamic viscosities of 1,2,4-trichlorobenzene (1) + $n$-alkane (2) mixtures at 293.15 K and atmospheric pressure. Points, experimental results [15]: (●), decane; (■), tetradecane. Solid lines, results from the application of the Bloomfield-Dewan´s model (equation 5) with ($\alpha =1$, $\beta = 0$)

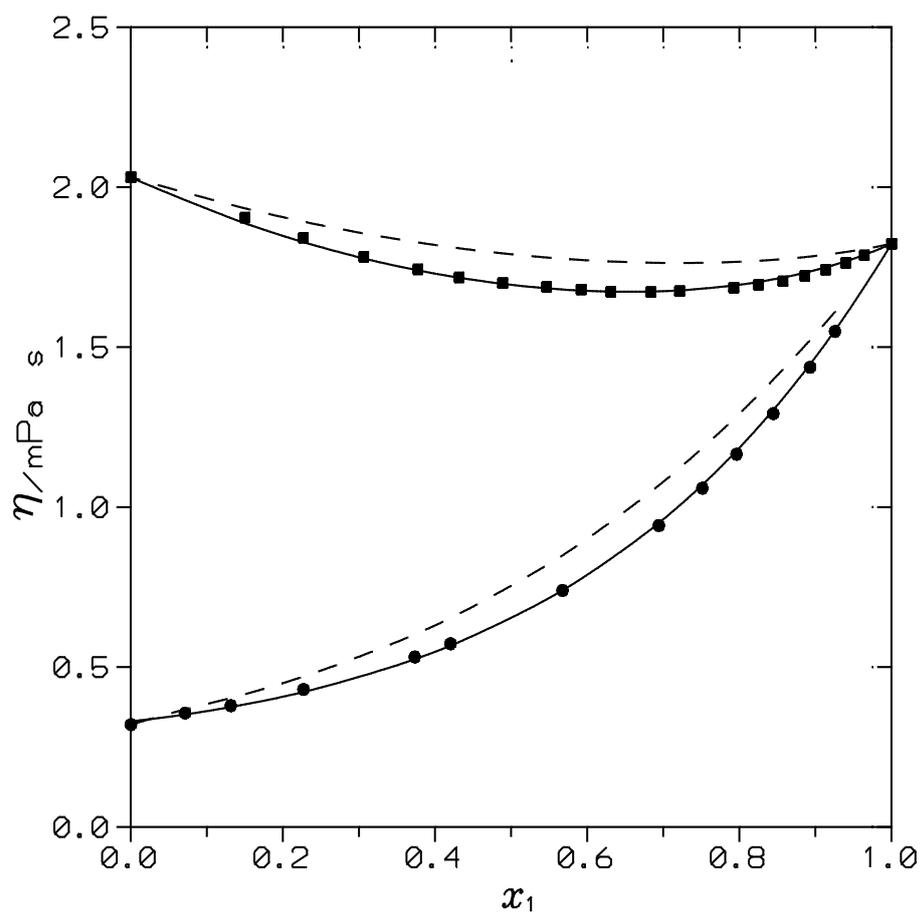

**Figure 6.**  Dynamic viscosities of methyl benzoate (1) + $n$-alkane (2) mixtures at 298.15 K and atmospheric pressure. Points, experimental results [40]: (●), hexane; (■), tetradecane. Dashed lines, results from the application of the Bloomfield-Dewan´s model (equation 5) with ($\alpha =1$, $\beta = 0$); solid lines, results using ($\alpha =1$, $\beta = 0.37$) for the mixture with hexane and ($\alpha =1$, $\beta = 0.1$) for the solution with tetradecane.

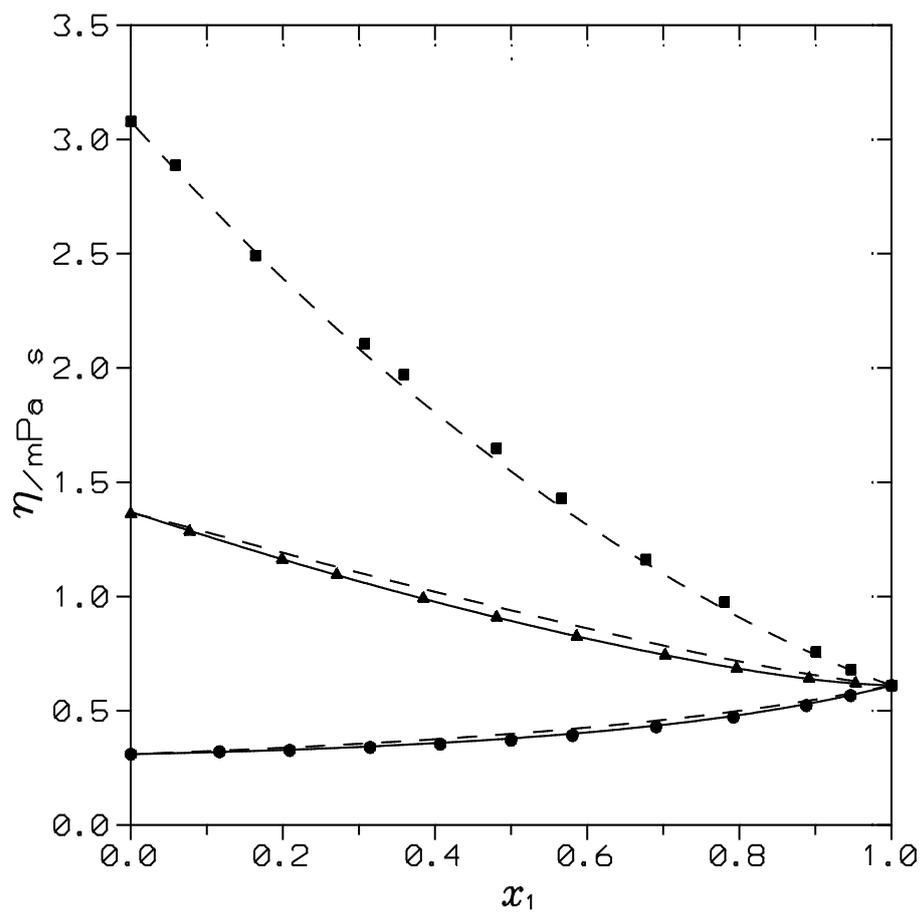

**Figure 7.** Dynamic viscosities of benzene (1) + *n*-alkane (2) mixtures at 298.15 K and atmospheric pressure. Points, experimental results [59]: (●), hexane; (▲), dodecane; (■), hexadecane. Dashed lines, results from the application of the Bloomfield-Dewan´s model (equation 5) with ($\alpha =1$, $\beta = 0$); solid lines, results using ($\alpha =1$, $\beta = 0.3$) for the mixture with hexane and ($\alpha =1$, $\beta = 0.25$) for the solution with dodecane.

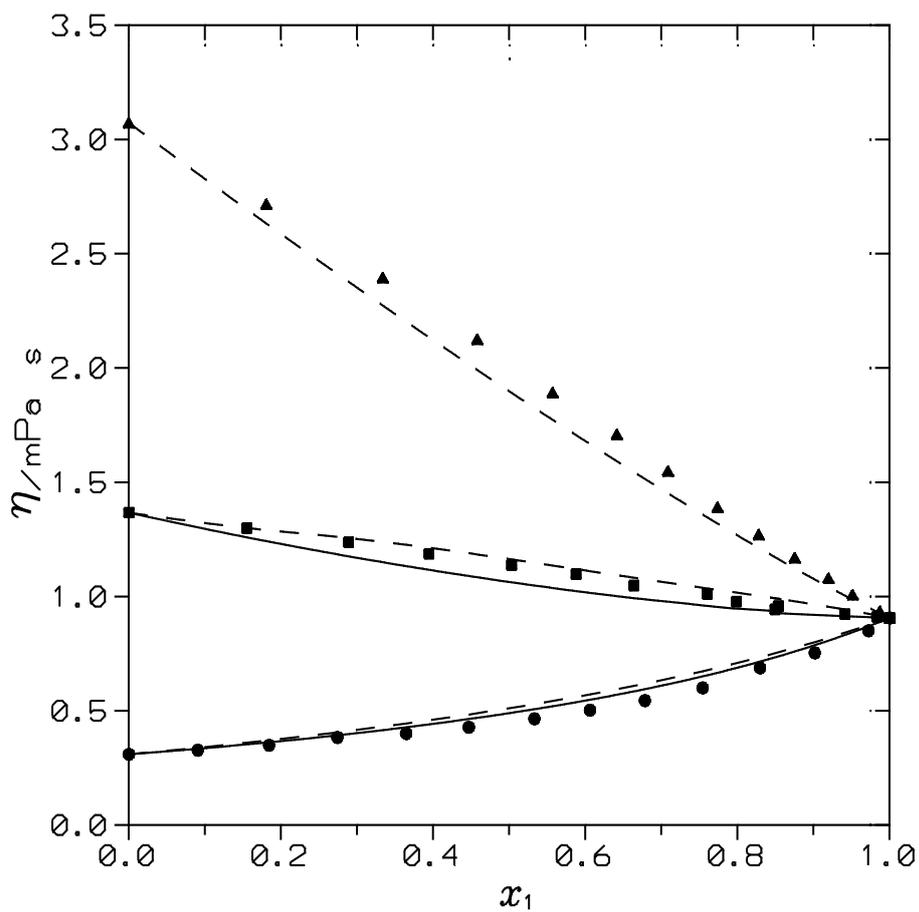

**Figure 8.** Dynamic viscosities of cyclohexane (1) + *n*-alkane (2) mixtures at 298.15 K and atmospheric pressure. Points, experimental results [60]: (●), hexane; (■), dodecane; (▲), hexadecane. Dashed lines, results from the application of the Bloomfield-Dewan´s model (equation 5) with ($\alpha = 1$, $\beta = 0$); solid lines, results using ($\alpha = 1$, $\beta = 1$) for the mixture with hexane and ($\alpha = 1$, $\beta = 0.7$) for the solution with dodecane.

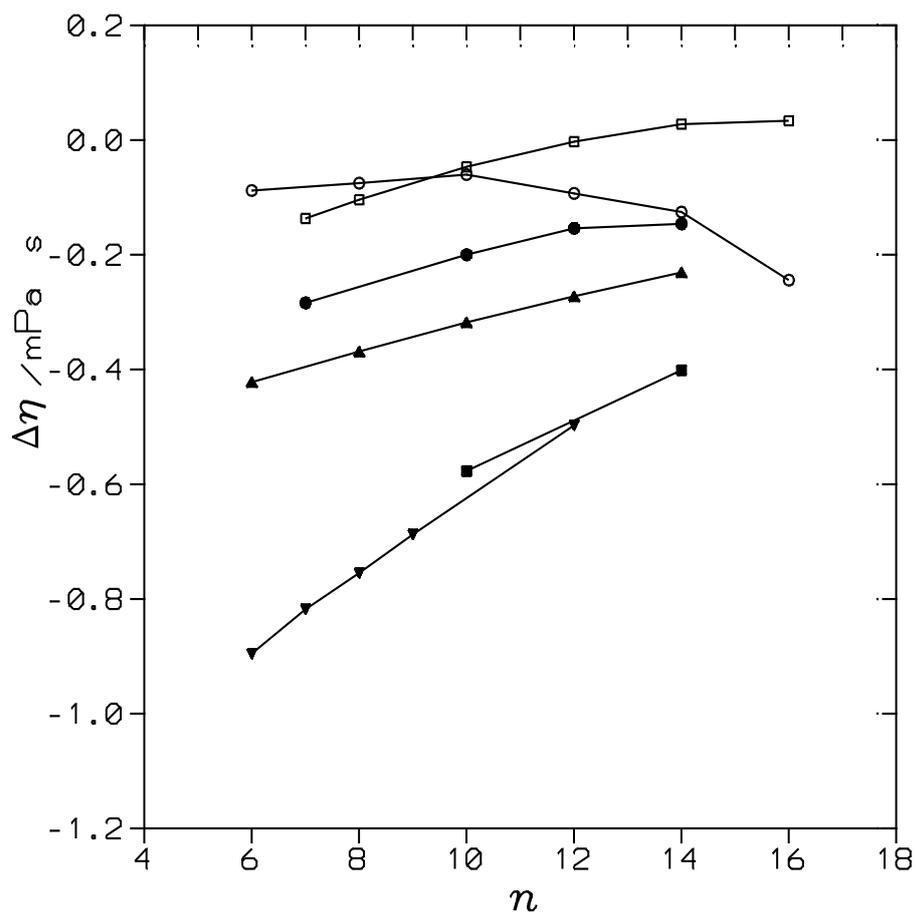

**Figure 9.** $\Delta\eta$ at 298.15 K, atmospheric pressure and equimolar composition for iodobenzene (1) (●), or 1,2,4-trichlorobenzene (1) (■, $T$ = 293.15 K); or 1-chloronapthalene (1) (▼); or methyl benzoate (1) (▲); or benzene (1) (O), or cyclohexane (1) (□) + $n$-alkane (2) mixtures. For references, see Table 6. Lines are for the aid of the eye.

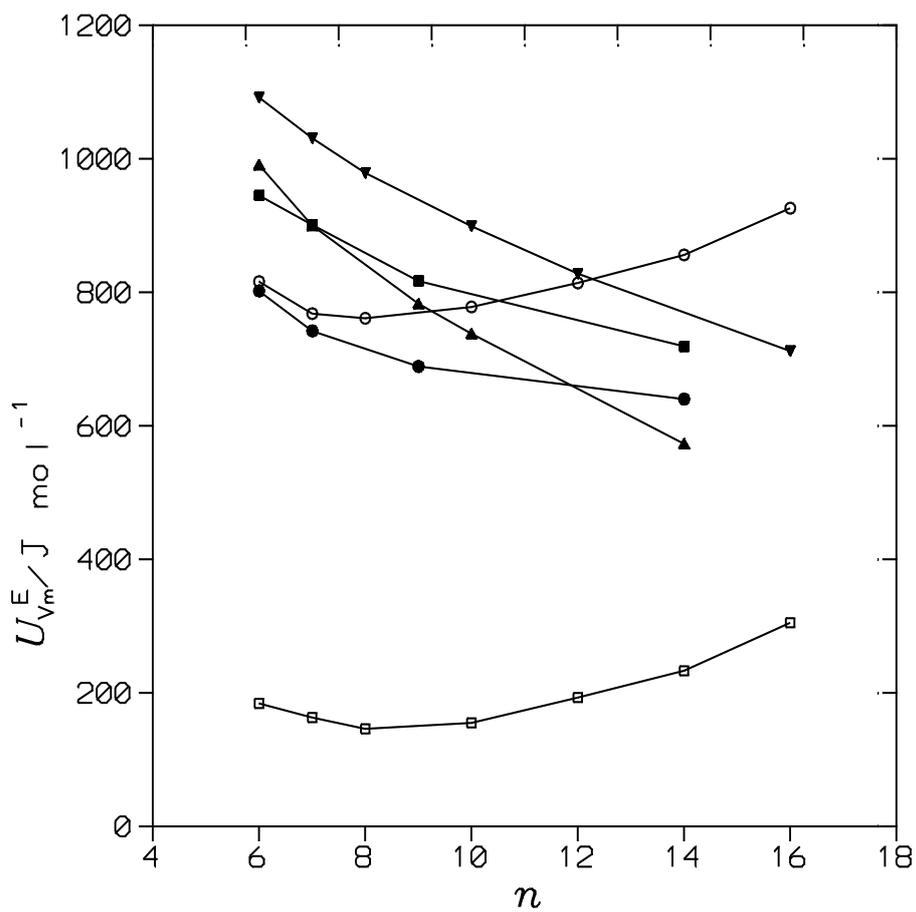

**Figure 10**   $U_{Vm}^E$ at 298.15 K, atmospheric pressure and equimolar composition and atmospheric pressure for chlorobenzene (1) (●), or bromobenzene (1) (■), or 1,2,4-trichlorobenzene (1) (▲), or 1-chloronapthalene (1) (▼), or benzene (1) (O), or cyclohexane (1) (□). + $n$-alkane (2) mixtures [10,11]. Lines are for the aid of the eye.

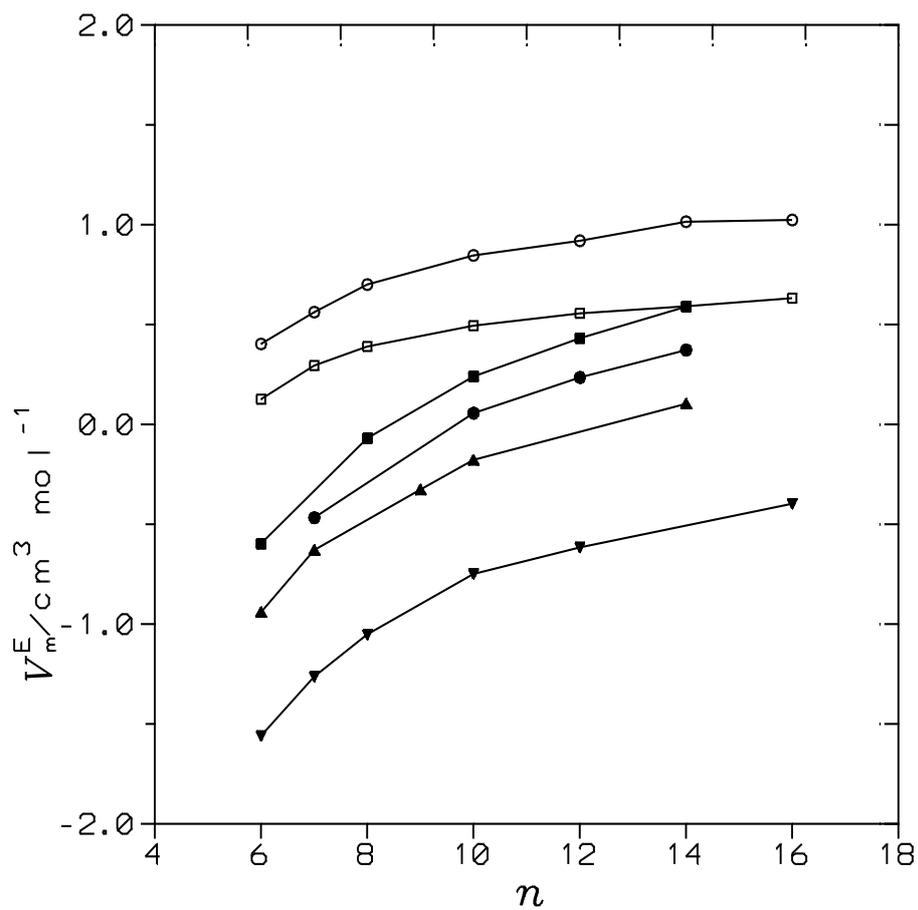

**Figure 11.** $V_m^E$ at 298.15 K, atmospheric pressure and equimolar composition for; iodobenzene (1) (●); or 1,2,4-trichlorobenzene (1) (▲); or 1-chloronapthalene (1) (▼); or methyl benzoate (1) (■); or benzene (1) (O); or cyclohexane (1) (□) + $n$-alkane (2) mixtures. For references, see Table 6. Lines are for the aid of the eye.

SUPPLEMENTARY MATERIAL

VISCOSITIES OF IODOBENZENE + *n*-ALKANE MIXTURES AT (288.15-308.15) K.
MEASUREMENTS AND RESULTS FROM MODELS

Luis Felipe Sanz[a], Juan Antonio González,[a*] Fernando Hevia[a], Daniel Lozano-Martín[a], Isaías García de la Fuente,[a] José Carlos Cobos[a]

## TABLE S1

Magnitudes of activation of viscous flow at 298.15 K and atmospheric pressure determined according to the Eyring's theory: enthalpy, $\Delta H_m^*$, entropy, $\Delta S_m^*$, Gibbs energy, $\Delta G_m^*$, and $\Delta(\Delta G_m^*)(= \Delta G_m^* - x_1 \Delta G_{m1}^* - x_2 \Delta G_{m2}^*)$ for the mixtures iodobenzene (1) + n-alkane (2).

| $x_1$ | $\Delta H_m^*$/(kJ mol⁻¹) | $\Delta S_m^*$/(J K⁻¹ mol⁻¹) | $\Delta G_m^*$/(kJ mol⁻¹) | $\Delta(\Delta G_m^*)$/(J mol⁻¹) |
|---|---|---|---|---|
| \multicolumn{5}{c}{iodobenzene (1) + heptane (2)} ||||
| 0. | 7.2 | − 17.1 | 12.3 | |
| 0.0504 | 7.3 | − 17.0 | 12.4 | − 45 |
| 0.0951 | 7.4 | − 17.1 | 12.5 | − 97 |
| 0.1454 | 7.4 | − 17.3 | 12.6 | − 134 |
| 0.1925 | 7.5 | − 17.3 | 12.7 | − 181 |
| 0.2441 | 7.6 | − 17.4 | 12.8 | − 216 |
| 0.2929 | 7.6 | − 17.8 | 12.9 | − 243 |
| 0.3940 | 7.7 | − 18.1 | 13.1 | − 284 |
| 0.4954 | 8.3 | − 17.1 | 13.4 | − 303 |
| 0.5916 | 8.4 | − 17.6 | 13.6 | − 307 |
| 0.6936 | 9.1 | − 16.3 | 13.9 | − 278 |
| 0.7452 | 9.0 | − 17.2 | 14.1 | − 254 |
| 0.7958 | 9.3 | − 16.8 | 14.3 | − 205 |
| 0.8450 | 9.3 | − 17.3 | 14.4 | − 182 |
| 0.8963 | 9.7 | − 16.5 | 14.6 | − 128 |
| 0.9480 | 10.0 | − 16.2 | 14.8 | − 76 |
| 1. | 10.5 | − 15.4 | 15.0 | |
| \multicolumn{5}{c}{iodobenzene (1) + decane (2)} ||||
| 0 | 10.1 | − 16.5 | 15.0 | |
| 0.0968 | 9.8 | − 17.0 | 14.9 | − 71 |
| 0.1484 | 9.7 | − 17.2 | 14.9 | − 107 |
| 0.1980 | 9.7 | − 17.4 | 14.8 | − 140 |
| 0.2594 | 9.6 | − 17.7 | 14.8 | − 169 |
| 0.2959 | 9.6 | − 17.6 | 14.8 | − 188 |
| 0.3454 | 9.5 | − 17.8 | 14.8 | − 207 |
| 0.3939 | 9.5 | − 17.8 | 14.8 | − 226 |
| 0.4946 | 9.4 | − 18.0 | 14.8 | − 246 |
| 0.5945 | 9.4 | − 17.9 | 14.8 | − 253 |
| 0.6969 | 9.4 | − 18.2 | 14.8 | − 238 |
| 0.7461 | 9.4 | − 18.0 | 14.8 | − 224 |
| 0.7960 | 9.5 | − 17.8 | 14.8 | − 197 |
| 0.8465 | 9.3 | − 18.6 | 14.9 | − 161 |

TABLE S1 (continued)

| | | | | |
|---|---|---|---|---|
| 0.8974 | 9.6 | −17.9 | 14.9 | −127 |
| 0.9486 | 10.0 | −16.8 | 15.0 | −72 |
| 1 | 10.4 | 15.7 | 15.1 | |
| iodobenzene (1) + dodecane (2) | | | | |
| 0 | 12.3 | −14.1 | 16.5 | |
| 0.0988 | 11.9 | −14.8 | 16.3 | −34 |
| 0.1496 | 11.7 | −15.3 | 16.3 | −45 |
| 0.1964 | 11.6 | −15.5 | 16.2 | −63 |
| 0.2466 | 11.4 | −15.8 | 16.1 | −73 |
| 0.2971 | 11.2 | −16.1 | 16.0 | −88 |
| 0.3978 | 10.8 | −16.7 | 15.8 | −104 |
| 0.4966 | 10.7 | −16.7 | 15.7 | −122 |
| 0.5964 | 10.6 | −16.6 | 15.5 | −129 |
| 0.6974 | 10.2 | −17.3 | 15.4 | −123 |
| 0.7480 | 10.1 | −17.3 | 15.3 | −119 |
| 0.7977 | 10.1 | −17.3 | 15.2 | −106 |
| 0.8475 | 9.9 | −17.6 | 15.2 | −99 |
| 0.8982 | 9.7 | −18.4 | 15.2 | −74 |
| 0.9487 | 10.1 | −16.7 | 15.1 | −44 |
| 1 | 10.4 | −15.6 | 15.1 | |
| iodobenzene (1) + tetradecane (2) | | | | |
| 0 | 14.5 | −11.4 | 17.9 | |
| 0.1007 | 13.9 | −12.5 | 17.6 | 21 |
| 0.1478 | 13.6 | −13.0 | 17.5 | 28 |
| 0.1966 | 13.4 | −13.5 | 17.4 | 34 |
| 0.2479 | 13.2 | −14.0 | 17.2 | 42 |
| 0.2885 | 12.8 | −14.4 | 17.1 | 44 |
| 0.3485 | 12.5 | −14.8 | 17.0 | 48 |
| 0.3972 | 12.3 | −15.2 | 16.8 | 50 |
| 0.4991 | 11.8 | −16.0 | 16.5 | 49 |
| 0.5981 | 11.4 | −16.4 | 16.2 | 44 |
| 0.6981 | 11.0 | −16.6 | 15.9 | 28 |
| 0.7483 | 10.8 | −16.7 | 15.8 | 20 |
| 0.7982 | 10.7 | −16.7 | 15.6 | 12 |
| 0.8492 | 10.4 | −17.1 | 15.5 | 5 |
| 0.8986 | 10.5 | −16.3 | 15.3 | 1 |
| 0.9490 | 10.3 | −16.4 | 15.2 | −8 |
| 1 | 10.3 | −15.8 | 15.1 | |

## TABLE S2

Results from the application of the correlation equations Grunberg-Nissan (eq. 3), and Fang-He (eq. 4) to iodobenzene (1), or 1-chloronaphthalene (1), or 1,2,4-trichlorobenzene (1), or methyl benzoate (1), or benzene (1) or + cyclohexane (1) + $n$-alkane (2) mixtures at 298.15 K and atmospheric pressure.

| System[a] | $N$[b] | $G_{12}$[c] | $\sigma_r(\eta)$[d] | $W_{12}/(RT)$[e] | $\sigma_r(\eta)$[d] |
|---|---|---|---|---|---|
| iodobenzene + $n$-C$_7$ | 15 | −0.593 | 0.006 | −0.006 | 0.014 |
| iodobenzene + $n$-C$_{10}$ | 15 | −0.634 | 0.018 | −0.010 | 0.006 |
| iodobenzene + $n$-C$_{12}$ | 14 | −0.510 | 0.020 | −0.010 | 0.007 |
| iodobenzene + $n$-C$_{14}$ | 15 | −0.310 | 0.030 | −0.010 | 0.008 |
| 1-chloronaphthalene + $n$-C$_6$ | 9 | −0.508 | 0.035 | −0.022 | 0.035 |
| 1-chloronaphthalene + $n$-C$_7$ | 9 | −0.637 | 0.018 | −0.020 | 0.018 |
| 1-chloronaphthalene + $n$-C$_8$ | 9 | −0.693 | 0.020 | −0.017 | 0.018 |
| 1-chloronaphthalene + $n$-C$_{10}$ | 9 | −0.711 | 0.029 | −0.013 | 0.030 |
| 1-chloronaphthalene + $n$-C$_{12}$ | 9 | −0.664 | 0.030 | −0.010 | 0.022 |
| 1,2,4-trichlorobenzene + $n$-C$_{10}$[f] | 9 | −0.526 | 0.014 | −0.008 | 0.009 |
| 1,2,4-trichlorobenzene + $n$-C$_{14}$[f] | 9 | −0.193 | 0.003 | −0.005 | 0.010 |
| methyl benzoate + $n$-C$_6$ | 12 | −0.682 | 0.008 | −0.025 | 0.009 |
| methyl benzoate + $n$-C$_8$ | 17 | −0.769 | 0.004 | −0.020 | 0.006 |
| methyl benzoate + $n$-C$_{10}$ | 18 | −0.801 | 0.007 | −0.019 | 0.007 |
| methyl benzoate + $n$-C$_{12}$ | 16 | −0.709 | 0.015 | −0.017 | 0.014 |
| methyl benzoate + $n$-C$_{14}$ | 18 | −0.504 | 0.013 | −0.014 | 0.004 |
| cyclohexane + $n$-C$_6$ | 12 | −0.467 | 0.040 | −0.013 | 0.039 |
| cyclohexane + $n$-C$_7$ | 12 | −0.616 | 0.020 | −0.017 | 0.016 |
| cyclohexane + $n$-C$_8$ | 12 | −0.478 | 0.026 | −0.011 | 0.025 |
| cyclohexane + $n$-C$_{10}$ | 12 | −0.217 | 0.018 | −0.004 | 0.015 |
| cyclohexane + $n$-C$_{12}$ | 12 | 0.074 | 0.013 | −1.5 10$^{-5}$ | 0.014 |
| cyclohexane + $n$-C$_{14}$ | 12 | 0.413 | 0.039 | 0.004 | 0.036 |
| cyclohexane + $n$-C$_{16}$ | 12 | 0.771 | 0.015 | 0.007 | 0.007 |
| benzene + $n$-C$_6$ | 10 | −0.640 | 0.005 | −0.018 | 0.007 |
| benzene + $n$-C$_8$ | 9 | −0.590 | 0.009 | −0.017 | 0.011 |
| benzene + $n$-C$_{10}$ | 10 | −0.300 | 0.012 | −0.013 | 0.015 |
| benzene + $n$-C$_{12}$ | 10 | −0.060 | 0.012 | −0.014 | 0.002 |
| benzene + $n$-C$_{14}$ | 9 | 0.282 | 0.014 | −0.015 | 0.023 |

TABLE S2 (continued)

| | | | | | |
|---|---|---|---|---|---|
| benzene + $n$-C$_{16}$ | 10 | 0.560 | 0.020 | −0.011 | 0.021 |

[a]for references of experimental data, see Table 6; [b] number of data points; [c]adjustable parameter in the Grunberg-Nissan equation (eq.3); [d]standard relative deviation (eq. 12); [e]adjustable parameter in the Fang-He equation (eq. 4); [e]system at 293.15 K.

## TABLE S3

Magnitudes of activation of viscous flow determined according to the Eyring's theory: Gibbs energy, $\Delta G_m^*$, and $\Delta(\Delta G_m^*)(=\Delta G_m^* - x_1\Delta G_{m1}^* - x_2\Delta G_{m2}^*)$ for aromatic compound (1) or cyclohexane (1) + $n$-alkane (2) mixtures at 298.15 K, atmospheric pressure and equimolar composition.

| System | $\Delta G_m^*$/(kJ mol$^{-1}$) | $\Delta(\Delta G_m^*)$/(J mol$^{-1}$) | Ref. |
|---|---|---|---|
| iodobenzene (1) + $n$-C$_7$ (2) | 13.3 | −309 | this work |
| iodobenzene (1) + $n$-C$_{10}$ (2) | 14.7 | −248 | this work |
| iodobenzene (1) + $n$-C$_{12}$ (2) | 15.6 | −120 | this work |
| iodobenzene (1) + $n$-C$_{14}$ (2) | 16.6 | 49 | this work |
| 1-chloronaphthalene (1) + $n$-C$_6$ (2) | 13.8 | −380 | 16 |
| 1-chloronaphthalene (1) + $n$-C$_7$ (2) | 14.4 | −412 | 16 |
| 1-chloronaphthalene (1) + $n$-C$_8$ (2) | 14.8 | −425 | 16 |
| 1-chloronaphthalene (1) + $n$-C$_{10}$ (2) | 15.7 | −410 | 16 |
| 1-chloronaphthalene (1) + $n$-C$_{12}$ (2) | 16.4 | −343 | 16 |
| methyl benzoate (1) + $n$-C$_6$ (2) | 13.2 | −428 | 40 |
| methyl benzoate (1) + $n$-C$_8$ (2) | 14.1 | −456 | 40 |
| methyl benzoate (1) + $n$-C$_{10}$ (2) | 14.9 | −435 | 40 |
| methyl benzoate (1) + $n$-C$_{12}$ (2) | 15.8 | −324 | 40 |
| methyl benzoate (1) + $n$-C$_{14}$ (2) | 16.6 | −143 | 40 |
| benzene (1) + $n$-C$_6$ (2) | 11.5 | −328 | 59 |
| benzene (1) + $n$-C$_8$ (2) | 12.5 | −238 | 59 |
| benzene (1) + $n$-C$_{10}$ (2) | 13.6 | 32 | 59 |
| benzene (1) + $n$-C$_{12}$ (2) | 14.6 | 245 | 59 |
| benzene (1) + $n$-C$_{14}$ (2) | 15.5 | 452 | 59 |
| benzene (1) + $n$-C$_{16}$ (2) | 16.5 | 828 | 59 |
| cyclohexane (1) + $n$-C$_6$ (2) | 12.3 | −260 | 60 |
| cyclohexane (1) + $n$-C$_7$ (2) | 12.7 | −327 | 60 |
| cyclohexane (1) + $n$-C$_8$ (2) | 13.2 | −270 | 60 |
| cyclohexane (1) + $n$-C$_{10}$ (2) | 14.3 | 0 | 60 |
| cyclohexane (1) + $n$-C$_{12}$ (2) | 15.3 | 213 | 60 |
| cyclohexane (1) + $n$-C$_{14}$ (2) | 16.3 | 520 | 60 |
| cyclohexane (1) + $n$-C$_{16}$ (2) | 17.2 | 796 | 60 |

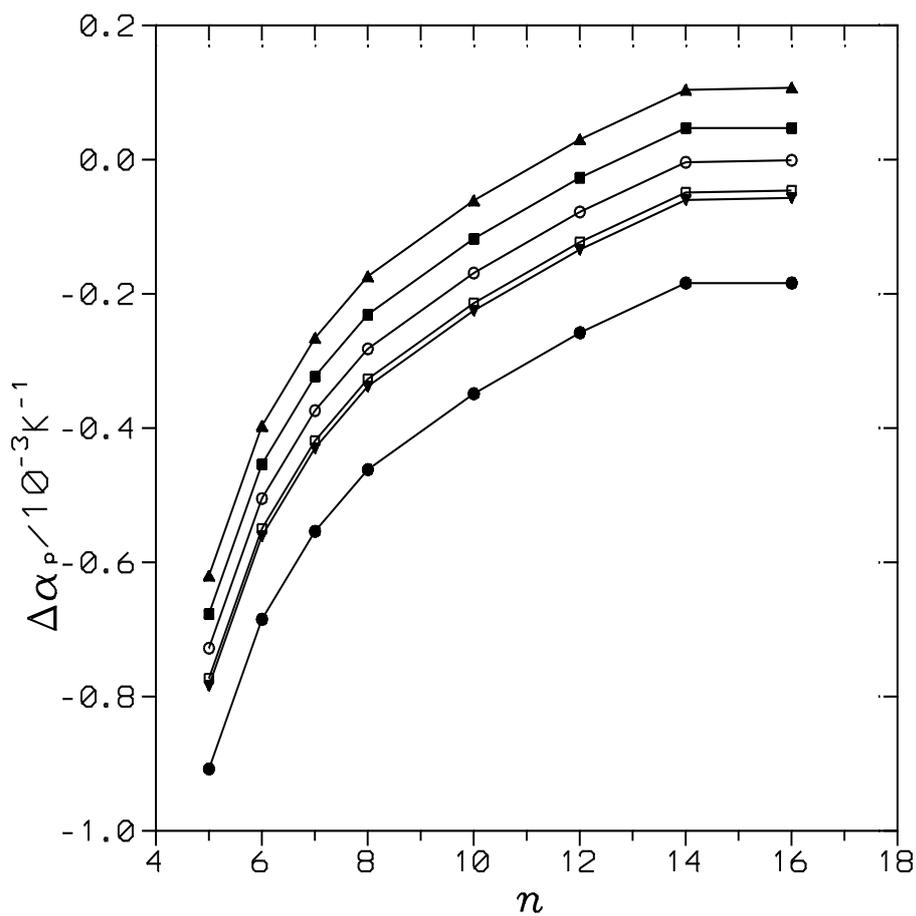

**Figure S1.** Differences between thermal expansion coefficients, $\Delta\alpha_p = (\alpha_{p1} - \alpha_{p2})$, of mixture compounds in aromatic compound(1) + $n$-alkane(2) systems at 298.15 K and atmospheric pressure vs. $n$, the number of C atoms of the $n$-alkane: (●), 1-chloronaphthalene; (▼), 1,2,4-trichlorobenzene; (□), iodobenzene; (O), methyl benzoate; (■), bromobenzene; (▲), chlorobenzene (for references, see [10-12,40,74]. Lines are for the aid of the eye.

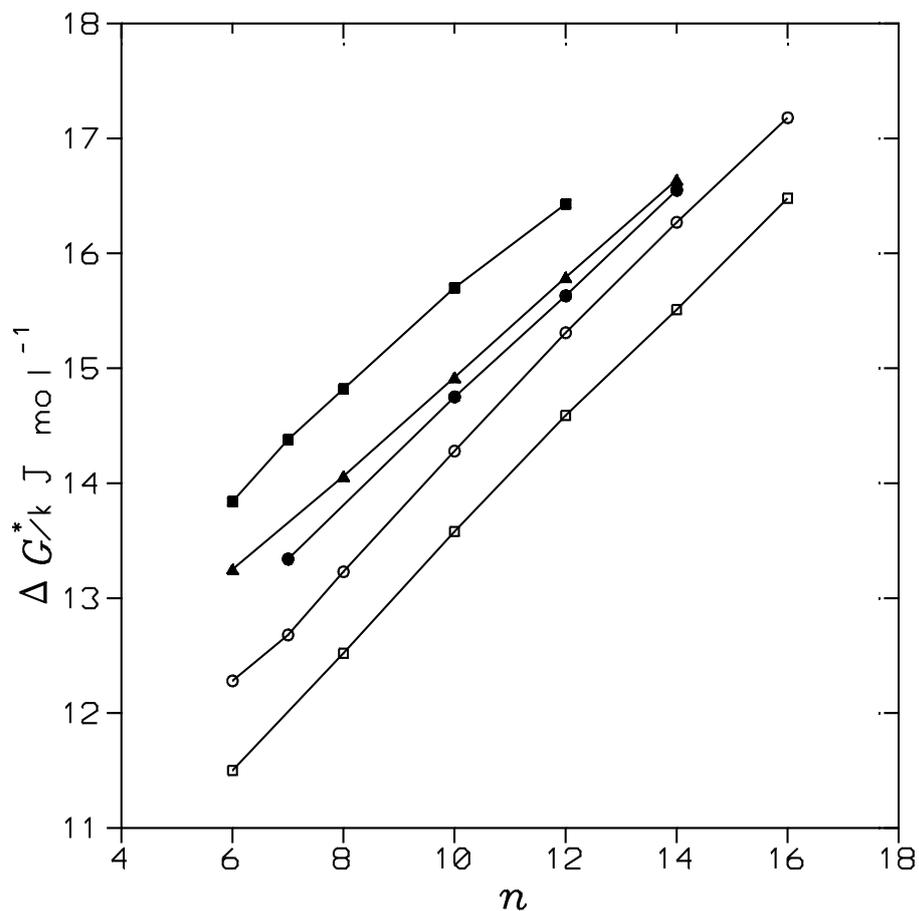

**Figure S2.** Values of $\Delta G_m^*$ at 298.15 K, atmospheric pressure, and equimolar composition for iodobenzene (1) (●), or 1-chloronaphthalene (1) (■), or methyl benzoate (1) (▲), or cyclohexane (1) (O), or benzene (1) (□) + *n*-alkane (2) mixtures. Lines are for the aid of the eye.